\documentclass[journal, twoside]{IEEEtran}
\usepackage{cite}

\ifCLASSINFOpdf
  \usepackage[pdftex]{graphicx}
\else
\fi
\usepackage{amsmath}
\usepackage{amssymb}
\usepackage{array}

\ifCLASSOPTIONcompsoc
 \usepackage[caption=false,font=normalsize,labelfont=sf,textfont=sf]{subfig}
\else
 \usepackage[caption=false,font=footnotesize]{subfig}
\fi
\usepackage{fixltx2e}

\usepackage{stfloats}
\ifCLASSOPTIONcaptionsoff
 \usepackage[nomarkers]{endfloat}
\let\MYoriglatexcaption\caption
\renewcommand{\caption}[2][\relax]{\MYoriglatexcaption[#2]{#2}}
\fi
\let\MYorigsubfloat\subfloat
\renewcommand{\subfloat}[2][\relax]{\MYorigsubfloat[]{#2}}
\usepackage{hyperref}
\hypersetup{
    colorlinks=true,
    linkcolor=blue,
    filecolor=magenta,      
    urlcolor=black,
}

\usepackage{xcolor}
\begin{document}
%
\title{RE-MIMO: Recurrent and Permutation Equivariant \\ Neural MIMO Detection}
%
%
%

\author{Kumar~Pratik,~Bhaskar~D.~Rao,~and~Max~Welling
\thanks{Kumar Pratik and Max Welling was with the Informatics Institute, University of Amsterdam, Science Park 904, The Netherlands. (e-mail: kumar.pratik73@yahoo.com, m.welling@uva.nl).}
\thanks{B. D. Rao is with the Electrical and Computer Engineering Department,
University of California, San Diego, CA 92093, USA. (e-mail: brao@ucsd.edu). B. Rao's research was supported in part by ONR Grant No. N00014-18-1-2038}
\thanks{Manuscript received June 30, 2020; revised June 30, 2020.}}

%
%

\markboth{IEEE TRANSACTIONS ON SIGNAL PROCESSING,~Oct~2020}%
{Pratik,~Rao,~and~Welling: RE-MIMO- Recurrent and Permutation Equivariant Neural MIMO Detection}
%



\maketitle

\begin{abstract}
    In this paper, we present a novel neural network architecture for MIMO symbol detection, the \textit{Recurrent Equivariant MIMO detector} (RE-MIMO). It incorporates several important considerations in wireless communication systems, such as robustness to channel misspecification, the ability to handle a varying number of users with a single model, and invariance to the (sequential) order in which the users interact with the system. The decoder consists of three main blocks; one block to inform the decoder of the channel model, one permutation equivariant block based on the successful transformer architecture to model transmitter channel interactions, and one fully connected feedforward network to predict the demodulated symbols for each user. These blocks are chained together into an iterative decoder that is trained through end-to-end backpropagation.  
    RE-MIMO is compared against a broad range of existing methods and the results confirm the ability of the network to achieve the state of the art demodulation accuracy. In particular, RE-MIMO efficiently handles a variable number of transmitters with only a single trained model, is capable of dealing with correlated channels, and is robust to channel misspecification. In terms of computational complexity, RE-MIMO scales favorably with the number of transmitters.

\end{abstract}

\begin{IEEEkeywords}
Massive MIMO, Receiver Design, Deep Learning, Recurrent inference, Attention networks
\end{IEEEkeywords}

%
\IEEEpeerreviewmaketitle

\section{Introduction}
\label{sec:intro}
    \IEEEPARstart{M}{ultiple} input multiple output (MIMO) systems play a very important role in improving the data throughput in modern wireless communication systems. This comes at the expense of increased receiver complexity. A range of receivers has been developed that offer a trade-off between complexity and performance \cite{paulraj2004overview,yang2015fifty,blast,pic_1, pic_2,sdr_relax, sdr_qam,amp,oamp}. Simple linear receivers such as zero-forcing and minimum mean squared error detectors perform poorly. Efficient and effective symbol detection methods such as the sphere decoder~\cite{vikalo2004iterative,1603705} have been developed and are suitable only for small MIMO systems due to complexity reasons.
    
    Massive MIMO being considered for the next-generation wireless systems, exacerbate the complexity problem \cite{larsson2014massive,marzetta2016fundamentals,chockalingam2014large}. The spatial degrees of freedom offered by the massive MIMO systems enable supporting many users ($N_{tr}$) simultaneously. The desire for high data rates leads to the use of higher modulation ($M$) rates. The combination of multiple users and higher data rates makes the optimal maximum likelihood detector impractical due to the exponential complexity ($O(M^{N_{tr}})$) of the search space. Hence, a low complexity detection algorithm that scales to massive MIMO systems with higher modulation schemes has been an active topic of research. Several detectors have been designed for this purpose such as those based on local search, probabilistic data association, message passing, etc \cite{chockalingam2014large}. Algorithms based on message passing reduce complexity by utilizing the channel hardening property that is only likely to occur for large systems with i.i.d. fading channels.
    
    Recent developments in deep learning (DL)~\cite{dl} based architectures have shown considerable promise in many applications, e.g. object recognition, language modeling, etc. They offer an interesting opportunity to revisit the massive MIMO receiver design problem and explore the complexity versus performance trade-off. Several learning-based MIMO detection schemes~\cite{deepmimo, oampnet, oamp_net_2, mmnet, dl_mimo_1, dl_mimo_2} have been proposed in the recent past and they show considerable potential. However, there is much to be done in terms of identifying appropriate architectures suitable for wireless communication and demonstrating their efficacy. The neural network architecture developed in this paper incorporates several novel elements to address issues pertinent to communications. Though the work is limited to linear models currently in use for massive MIMO systems, the insights gained should prove useful to future systems that incorporate non-linearity with bigger potential payoffs.
    
    In this paper, we introduce \textit{Recurrent Equivariant MIMO detector} (RE-MIMO), the first Transformer~\cite{transformers} based Recurrent Inference Network (RIM)~\cite{RIM, rim_6} dedicated to wireless communication applications. The main guiding principles and key contributions of the neural detector design are incorporating permutation equivariance and dealing with varying number of transmitters. In MIMO systems, there is no natural ordering of the transmitters. Hence, any ordering imposed on the model will be artificial and provide opportunities for the training procedure to overfit on the (artificial) ordering observed in the training data. The goal of the network designed should be to remove any spurious signals in the training data that do not generalize to test data. We do so by making the model equivariant to permutations of the transmitters. This means that if we permute the order of the transmitters in the input, we get the equivalent permutation for the predictions at the output. Moreover, and maybe more importantly, we want the system to effortlessly handle a varying number of transmitters without having to use separate models for each situation with a different number of transmitters. The network we propose, based on the ideas behind graph neural networks~\cite{gnn_1, gnn_2, gnn_3}, transformers~\cite{transformers}, and deep sets~\cite{deepS} achieves that in a very natural way. Such permutation equivariant models can use the correct inductive bias, can handle a variable number of transmitters, and are more data efficient because they share parameters between instances with different numbers of transmitters.
    
    Another important consideration and novelty of the neural detector design is to explicitly incorporate the generative model (a.k.a. forward model) into the neural network. Most networks for solving linear inverse problems fall into one of two categories: either they explicitly invert the forward model $\mathbf{y}=\mathbf{Hx}+\mathbf{n}$ through least squared estimators or iterative estimators such as belief propagation~\cite{oampnet, oamp_net_2, sparse_amp}, or they collect a large number of data and then train a neural net to predict 
    $\hat{\mathbf{x}}=f(\mathbf{H}, \mathbf{y})$~\cite{FF1, FF2, FF3, FF4, FF5}. Both have disadvantages. The first one relies on the forward model being correct and on the accuracy of the inverse algorithm. The second is less constrained by the modeling assumptions (a.k.a. inductive biases) but now need to learn a possible complex relationship directly from data while ignoring known and potentially useful information about the forward model. We propose an intermediate strategy which we call "neural augmentation". In neural augmentation one trains a network from data but provides the network as much information about the generative, forward model as possible. This idea was proposed in MRI image reconstruction~\cite{rim_5}, dynamical systems modeling~\cite{rim_3}, and in error correction decoding~\cite{rim_4}, under the name RIM~\cite{RIM}.
    
    Another key feature of the neural architecture is the utilization of a transformer based attention network. Since its advent, transformers~\cite{transformers} have seen great success in language modeling, computer vision, deep reinforcement learning, etc. Their use for receiver design is apt because of its ability to model long-range dependencies, a feature important for dealing with multiple, potentially varying, number of transmitters.
    
    The outline of the paper is as follows. Section~\ref{sec:background} introduces  the problem definition and  provides a brief background on existing detection schemes. Section~\ref{sec:high_level} discusses the guiding principles of the RE-MIMO design and also includes a high level overview of RE-MIMO. Section~\ref{sec:RE-MIMO} presents an in-depth explanation of RE-MIMO architecture while section~\ref{sec:experiments} discusses the experimental findings. Finally, section~\ref{sec:conclusion} concludes our work with a brief conclusion and discussion about possible future directions.

\section{Background}
\label{sec:background}
    \subsection{Notation}
        The following notations are used in the paper. Lowercase symbols denote scalars, bold lowercase symbols denote vectors, and bold uppercase symbols denote matrices. $\mathbf{I}_n$ represents a complex identity matrix of size n. Unless stated otherwise, $x_i$ denotes the $i^{th}$ element of the vector $\mathbf{x}$. To differentiate between the ground truth transmitted signal and the estimated symbol for user $i$, we represent the estimated symbol (pre-softmax) by $\hat{\mathbf{x}}_{i}$ and the true symbol by $x_i$. Unless mentioned otherwise, the superscript on a vector denotes the iteration/time step, and the subscript refers to user index, i.e. $\hat{\mathbf{x}}^t_{i}$ is the prediction for user $i$ at time step $t$. The terms transmitters, users, and transmitter antennas mean the same thing, and henceforth will be used interchangeably throughout the paper.
\subsection{Problem definition}
    The forward model of a MIMO system is given by:
    \begin{equation*} \label{eq:1} \tag{1} 
        \mathbf{y}=\mathbf{Hx}+\mathbf{n}
    \end{equation*}
    where $\mathbf{H} \in \mathbb{C}^{N_{r} \times N_{tr}}$ is the channel matrix, $\mathbf{n} \sim \mathcal{C} \mathcal{N}\left(0, \sigma^{2} \mathbf{I}_{N_{r}}\right)$ is complex circular Gaussian noise, $\mathbf{x} \in \mathcal{X}^{N_{tr}}$ is the vector of transmitted symbols, $\mathbf{y} \in \mathbb{C}^{N_{r}}$ is the received signal vector, and $\mathcal{X}$ represents the finite set of constellation points. In our implementation, we assume a quadrature amplitude modulation (QAM)~\cite{qam} scheme, and constellation symbols are normalized to unit average power. We assume the same constellation set $\mathcal{X}$ for every transmitter, and every symbol in $\mathcal{X}$ has a uniform probability of being chosen by the transmitters. $N_{tr}$ denotes the number of single-antenna transmitters, and $N_r$ represents the number of receiver antennas located at the Base Station (BS). Perfect channel state information (CSI) is assumed, i.e., the channel matrix $\mathbf{H}$ and noise variance $\sigma^{2}$ are assumed to be known perfectly at the receiver. The  \textit{Maximum Likelihood} detector finds an estimate $\hat{\mathbf{x}}$ of $\mathbf{x}$, such that:
    \begin{equation*} \label{eq:2} \tag{2}            
        \hat{\mathbf{x}}=\underset{\mathbf{x} \in \mathcal{X}^{N_{tr}}}{\text{arg min }}\|\mathbf{y}-\mathbf{H x}\|_{2}^{2}
    \end{equation*}
    A search-based optimization algorithm (e.g., ML detector) that tries to find the configuration of transmitted symbols jointly, by minimizing the loss function mentioned above is NP-hard in complexity.
   \subsection{Classical MIMO detection schemes}
        The simplest of all the detectors are the linear receivers, such as zero forcing (ZF), matched filter (MF), and minimum mean squared error (MMSE) detectors \cite{paulraj2004overview,yang2015fifty,chockalingam2014large}. 
        Linear detectors are quite practical because of their lower complexity but their performance is not on par with the optimal detector (ML).
        
        Optimal detection can be implemented either by performing a brute search over the entire space of possible transmitted vectors simultaneously or by using efficient search algorithms. The sphere decoder (SD)~\cite{1603705} is one such efficient search algorithm that prunes the search space where $\|\mathbf{y}-\mathbf{H} \mathbf{x}\|_{2}^{2} > r$. However, it has been shown that even when the SD is used, the expected computational complexity is exponential and impractical for many applications. A lot of promising variations have been proposed in the literature to improve the SD, but their computational complexity renders them infeasible for massive MIMO systems.
        
        Semidefinite relaxation (SDR)~\cite{sdr_relax, sdr_qam}, proposed as a computationally efficient approximation for the ML detector, is based on a convex relaxation of the ML problem. Since the ML detector has high complexity because of the non-convex optimization involved, SDR attempts to approximate it as a convex (semidefinite) program that can be solved in polynomial time.
        
        Approximate message passing (AMP)~\cite{amp} is an iterative decoding algorithm that is asymptotically optimal in the large MIMO system limit with i.i.d. Gaussian channels. It builds on the idea of performing approximate inference in a graphical model representation of the MIMO system via approximate message passing. Owing to its lower complexity, AMP is quite easy to implement. Although AMP performs excellently in the case of i.i.d. Gaussian channels, its performance degrades significantly and is often unreliable for other channel matrices (e.g., ill-conditioned or spatially correlated). Orthogonal AMP (OAMP)~\cite{oamp} was introduced to relax the strong restrictions posed by AMP on channel matrices. OAMP is designed for unitary-invariant matrices~\cite{unitarily} which makes it suitable for a wider range of scenarios than AMP. AMP is faster than SDR in practice but is less robust than SDR.
        
        There are several other advanced detectors proposed in the literature (e.g., V-BLAST~\cite{blast}, PIC~\cite{pic_1, pic_2}) whose performances are highly dependent on system size ratio and their assumptions on channel matrices. Their performances are quite good as long as the assumptions are met and it gets significantly worse as we deviate from the idealized settings.

    \subsection{Learning based detection schemes}
        There are several ways of designing a neural architecture for detection purposes. At one extreme, one can deploy a stack of fully connected layers while at the other extreme, one can unfold an existing iterative algorithm~\cite{sparse_amp} whose optimal parameters are learned from the data. There is no clear winner as both the approaches have their pros \& cons. In the former case, the model has less bias and is very flexible but requires a large amount of data to be trained. It also does not incorporate domain knowledge into the architecture. And in the latter case, the model is very easy to train as it has very few trainable parameters but the model is less flexible and might inherit biases from its parent algorithm. It is easier to incorporate domain knowledge in model-based architectures.
        
        DetNet~\cite{deepmimo} is a deep learning network for MIMO detection and is based on the idea of projected gradient descent. DetNet matches the performance of SDR (standard baseline for i.i.d. Gaussian channels) at lower modulation schemes, such as BPSK and QAM4. Compared to AMP, DetNet is more robust to ill-conditioned channels. However, its performance degrades for higher modulation schemes.
        
        OAMPNet~\cite{oampnet} is a good example of a model-driven deep learning network for MIMO detection. The network is designed by unfolding the iterative OAMP algorithm and adding two learnable parameters per layer. The network is easily trained as the number of learnable parameters is just twice the number of layers. It has been shown to improve the performance of its parent OAMP algorithm. As OAMPNet is based on the OAMP algorithm, it also makes the strong assumption that the channel matrices are unitary-invariant. Similar to OAMP, OAMPNet requires computing a matrix pseudo inverse at each iteration step resulting in complexity higher than its alternatives like AMP. OAMPNet-2, proposed in~\cite{oamp_net_2}, introduces two additional learnable parameters per layer to the existing OAMPNet framework. The additional parameters are introduced to make the network adaptable to various channel environments and accommodate channel estimation errors.

        MMNet, proposed in~\cite{mmnet}, is another learning-based detection scheme inspired by iterative soft-thresholding algorithms~\cite{fista, amp}. MMNet introduces flexibility into the existing algorithm with trainable parameters optimized for each channel realization. The performance is almost at par with the OAMPNet for i.i.d. Gaussian channels. For correlated channels, MMNet needs to be trained online for each channel realization, which incurs significant latency in the algorithm. Also, the online training is not a desirable trait as we do not have control over the network during online training and one ill-conditioned channel realization can make the network drift in a Problem definition and a sub-optimal direction. Hence, we do not consider MMNet in our experiments. 
        
        All the networks discussed above are designed for a fixed number of transmitters.
        
\section{High level overview of the RE-MIMO architecture and design principles}
\label{sec:high_level}
    We first discuss some principles that guide our design based on the needs of the communication problem. Then we discuss the building blocks of the network that facilitate the implementation of the design guidelines.
    \subsection{Guiding principles for the RE-MIMO design}
         A crucial feature of RE-MIMO is user permutation equivariance. It enables the network to learn more efficiently, avoid over-fitting, and generalize better. A neural detector is a function $\mathbf{f}_{\theta}(\mathbf{H,y}):(\mathbf{H,y}) \rightarrow \hat{\mathbf{x}}$, which estimates the transmitted symbols, given $\mathbf{y}$ and $\mathbf{H}$. Permutation equivariance of the users implies that the ordering/labeling of the users should not impact network performance. It is relevant because reordering/relabeling the users simply permutes the columns of $\mathbf{H}$ along with an appropriate permutation of the entries of the symbol vector $\mathbf{x}.$ How the $\mathbf{H}$ is presented to the network should have no bearing on the neural detector. Instead of having the neural detector learn it from the data, it is significantly more beneficial from a learning and generalization point of view to have this as an inbuilt feature of the detector.
         
        \begin{figure}[ht]
            \centering
            \includegraphics[]{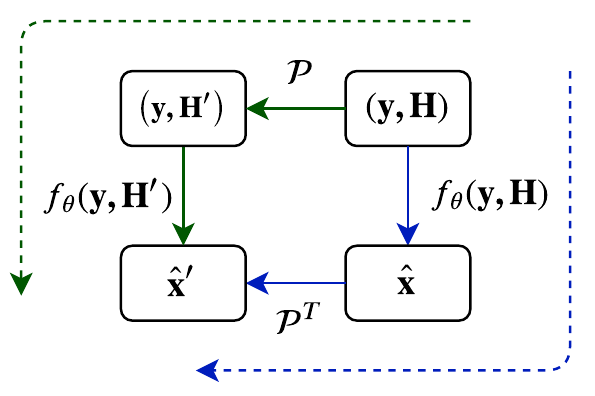}
            \caption{\textit{A flowchart depicting permutation equivariance property of neural RE-MIMO detector.} Irrespective of whether we permute the channel matrix and then decode the signal or decode the signal first and then permute the predicted symbols, we eventually arrive at the same result. Here, $\hat{\mathbf{x}}$ represents the reconstructed transmitted symbol vector.}
            \label{fig:perm_equi}
        \end{figure}
         More explicitly, let $\mathbf{\mathcal{P}}$ be a permutation matrix which when pre-multiplied or post-multiplied by a matrix, say $\mathbf{A}$, results in permuting the rows or columns of $\mathbf{A}$ respectively. Let $\mathbf{H}'$ represent the permuted channel matrix, where the columns are permuted according to $\mathbf{\mathcal{P}}$, i.e., $\mathbf{H}'=\mathbf{H \mathcal{P}}$. Now, let us permute the elements of the vector of transmitted symbols $\mathbf{x}$ by pre-multiplying it with $\mathbf{\mathcal{P}}^T$ to obtain $\mathbf{x}'$, i.e., $\mathbf{x}'=\mathbf{\mathcal{P}}^T\mathbf{x}$. The permuted $\mathbf{H}'$ and $\mathbf{x}'$ still result in the same received signal vector $\mathbf{y}$, which means $\mathbf{y}$ is invariant to the permutation of a given set of transmitter antennas. This permutation invariance property of $\mathbf{y}$ can be depicted as
        \begin{equation*} \label{eq:3} \tag{3}
            \mathbf{y} = \mathbf{H}'\mathbf{x}' + \mathbf{n} = \mathbf{H}\mathbf{\mathcal{P}} \mathbf{\mathcal{P}}^T\mathbf{x} + \mathbf{n} = \mathbf{H}\mathbf{x} + \mathbf{n}
        \end{equation*}
        The above formulation suggests that there is no natural ordering of elements in $\mathbf{x}$. Let $\hat{\mathbf{x}}'=\mathbf{f}_{\theta}(\mathbf{H',y})$ and $\hat{\mathbf{x}}=\mathbf{f}_{\theta}(\mathbf{H,y})$ represent the estimates of the transmitted symbols when function $\mathbf{f}_{\theta}$ is given $\mathbf{H}'$ and $\mathbf{H}$ respectively as the arguments. Then the permutation equivariance property implies $\hat{\mathbf{x}}'=\mathbf{\mathcal{P}}^T\hat{\mathbf{x}}$. The permutation equivariance property of the detector can mathematically be stated as
        \begin{equation*} \label{eq:4} \tag{4}
            \mathbf{\mathcal{P}}^{T} \mathbf{f}_{\theta}(\mathbf{H,y}) = \mathbf{f}_{\theta}(\mathbf{H \mathcal{P},y})
        \end{equation*}
        
         Another important feature of RE-MIMO is the ability to deal with a varying number of users. It is motivated by the dynamic nature of a communications system where the number of active users is continuously changing with network demand which is a function of time and user density. Training multiple networks with each network targeted to a specific user count is not practical and is severely limiting. Designing a network that can handle a varying number of users requires the network to be modular and share the same set of parameters for each active user. Moreover, from a communication system point of view, it must have an architecture that can handle the changing multi-user interference efficiently.
         
         Another significant feature of our network is the ability to incorporate domain knowledge. Given the linear model considered, there is a considerable structure that can be readily utilized. A combination of inductive bias and learning is appropriate in such scenarios and incorporated into RE-MIMO.

    \subsection{Neural Network architecture}
        \begin{figure*}[!t]
            \centering
            \includegraphics[]{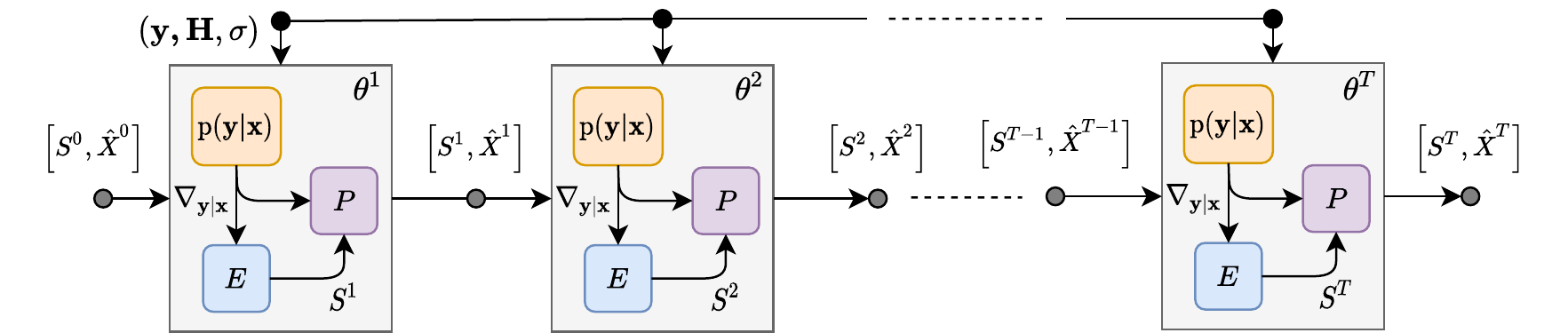}
            \caption{\textit{Illustration of the RE-MIMO framework.} RE-MIMO comprises of a stack of encoder-predictor ($EP$) blocks. Each gray shaded unit represents one $EP$ block, where $\theta^t$ denotes the set of trainable parameters in the $t^{th}$ $EP$ block. At every iteration, the encoder module ($E$) updates the previous state variables based on the information from the likelihood module $\text{p}(\mathbf{y|x})$ and the preceding $EP$ block. The predictor module then outputs the current signal reconstruction based on the updated state variables, previous signal reconstruction, and information from the likelihood module.}
            \label{fig:version1}
        \end{figure*}
        We present a high level overview of RE-MIMO that incorporates the design principles discussed above. The neural detector implements a recurrent estimation scheme as it learns an iterative decoding algorithm. The decoding scheme is unrolled in time as $T$ iteration steps. At each iteration, the detector produces a prediction (or reconstruction) of the transmitted symbols, which is expected to improve over time. The network also incorporates auxiliary state variables,  similar to standard RNNs. These state variables act as latent memory variables that store (abstract representations of) critical information required to reconstruct the transmitted symbols such as the amount of interference, number of co-transmitters involved, noise in the signal, or potentially more abstract concepts.
        
        The proposed RE-MIMO is implemented using a stack of iterative units as shown in Fig.~\ref{fig:version1}. The iterative unit is a neural computation module comprised of 3 sub-modules: the likelihood module, the encoder module, and the predictor module. The likelihood module injects information about the generative (forward) process into the neural network. The encoder-predictor (EP) modules together update the state vector and symbol estimates.  Let $S^t=\{\mathbf{s}^t_i\}_{i\in N_{tr}}$, $\hat{X}^t=\{\hat{\mathbf{x}}^t_i\}_{i\in N_{tr}}$ represent the sets of state variables and intermediate network predictions respectively. Each EP block takes as input previous state variables and previous predictions $[S^t, \hat{X}^t]$ from the preceding block, and outputs the updated state variables and current predictions $[S^{t+1}, \hat{X}^{t+1}]$. Each EP unit must be individually permutation equivariant and capable of handling a varying number of users $N_{tr}$.
        
        More specifically, the encoder module is primarily responsible for jointly updating the state variables $S_t=\{\mathbf{s}^t_{i}\}_{i \in N_{tr}}$ at every time step with the information required for making predictions. Hence, it must capture the dependencies between transmitters, i.e. each transmitter should be able to interact with every other transmitter in the encoder module irrespective of the number of users and be permutation equivariant. This is done with the incorporation of the transformer based self-attention network. Though there is coupling among all the states in the update process, it is useful to think of state $\mathbf{s}^t_i$  associated with user $i$ and the structure of the state update are designed to enable dealing with a variable number of users in a modular manner.
         
        The predictor is a transmitter specific point-wise classifier network purposed to produce a prediction individually for each of the users, based on its current state variable and previous reconstruction. It should be noted that $\hat{\mathbf{x}}^t_{i}$ does not refer to the numerical estimate of $x_i$ but instead represents the (pre-softmax) output vector of the predictor module, whose softmax, given by $\mathbf{p}(x_i|\textbf{y},\mathbf{H},\hat{\mathbf{x}}^{t}_{i})$, represents the probabilities of $x_i$ belonging to different symbols in $\mathcal{X}$.

\section{The Recurrent Equivariant Encoder and Predictor}
\label{sec:RE-MIMO}
    In this section, we discuss the technical details of the model components. We start by introducing the ideas behind the Recurrent Inference Machine (RIM)~\cite{RIM} which is the core of our RE-MIMO. Then we describe the details of the components in each iteration block: the likelihood module, the encoder module, and the predictor module.
    
    \subsection{Recurrent decoding}
        In this subsection, we begin with a brief introduction to the RIM~\cite{RIM} concept and then proceed to draw inspiration for RE-MIMO from the RIM point of view.
    
        The generative, forward model of a MIMO system is based on a linear measurement process given by (\ref{eq:1}). The symbol detection task can be accomplished using a maximum a posteriori (MAP) approach which reduces to an ML detector in case of a uniform prior on the symbols. In this work, we use a variant of MAP that has shown considerable promise in inverse problems such as image reconstruction~\cite{proximal} and medical tomography. The MAP solution involves solving the following optimization problem:
        \begin{equation*} \label{eq:5} \tag{5}
            \max_{\mathbf{x}} \log \text{p}(\mathbf{x}|\mathbf{y}) \propto \max_{\mathbf{x}} \left(\log \text{p}(\mathbf{y}|\mathbf{x}) + \log \text{p}_{\theta}(\mathbf{x})\right)
        \end{equation*}
        Here, $\log \text{p}(\mathbf{y}|\mathbf{x})$ is the log-likelihood term representing the underlying generative MIMO process (\ref{eq:1}) and $\text{p}_{\theta}(\mathbf{x})$ denotes a learnable, parametric prior over $\mathbf{x}$. Perhaps the simplest way to solve this optimization is through iterated gradient descent~\cite{grad}:
        \begin{equation*} \label{eq:6} \tag{6}
            \mathbf{x}^{t+1} = \mathbf{x}^{t} + \gamma_{t}\nabla\left(\log \text{p}(\mathbf{y|x}^t) + \log \text{p}_{\theta}(\mathbf{x}^t)\right)
        \end{equation*}
        where $\gamma_{t}$ denotes the step size at iteration $t$.
        
        \begin{figure*}[!t]
            \centering
            \includegraphics[]{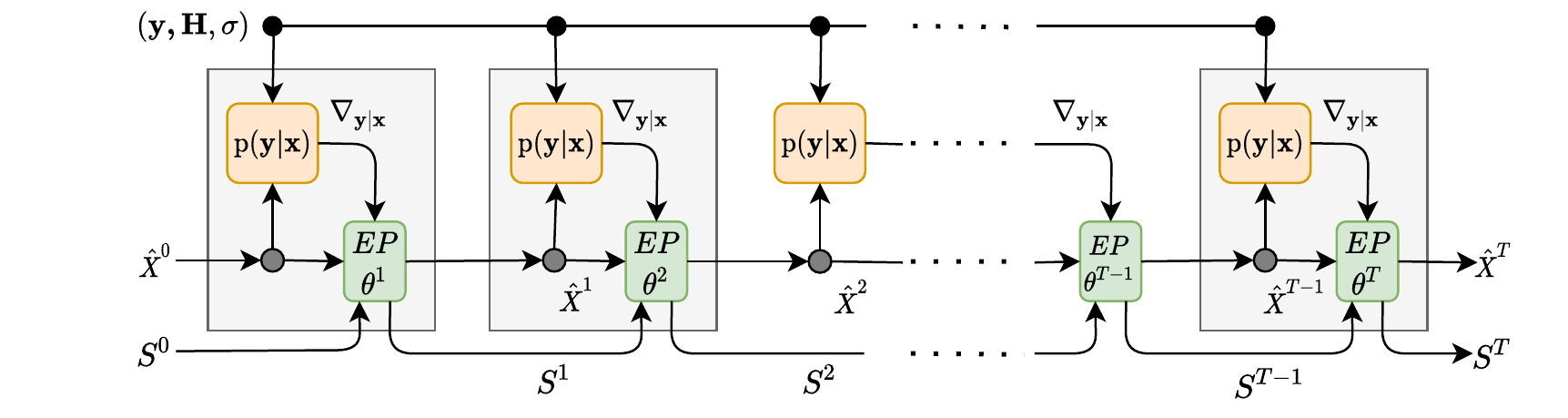}
            \caption{\textit{Graphical representation of the RE-MIMO rolled out in time.} The yellow and green boxes represent the likelihood $\text{p}(\mathbf{y|x})$ and the encoder-predictor ($EP$) modules respectively. $\theta^t$ is the set of trainable parameters associated with the $t^{th}$ $EP$ block. The detector aims to iteratively reconstruct the transmitted symbols from the received signal using the gradient of the likelihood model $\text{p}(\mathbf{y|x})$ and the previous reconstruction.}
            \label{fig:version2}
        \end{figure*}
        
        RIM is a general class of models that can learn an iterative inference algorithm without the need to explicitly specify a prior. Yet, it is more structured than simply learning a neural network on input-output pairs $\mathbf{\{x,y\}}$. It mimics the steps of an iterative inference/optimization process, but with general neural network blocks that receive information about the generative model to use in the estimation. As such, it sits between classical iterative decoders and full blown neural networks. The update equation can be written as:
        \begin{equation*}\label{eq:7} \tag{7}
            \mathbf{x}^{t+1} = \mathbf{x}^{t} + h_{\mu}^x\left(\nabla_{\mathbf{y|x}^t}, \mathbf{x}^t\right)
        \end{equation*}
        Here, $\nabla_{\mathbf{y|x}^t}$ denotes $\nabla \log \text{p}(\mathbf{y|x}^t)$, $h_{\mu}^x(.)$ is the neural update model for $\mathbf{x}$, and $\mu$ is the set of learnable parameters that govern the updates of $\mathbf{x}$. In the above generalization, the prior model parameters $\theta$ and step size parameters $\gamma$ have been merged into one set of trainable parameters $\mu$. The original update equation (\ref{eq:6}) can be obtained from
        \begin{equation*}\label{eq:8} \tag{8}
            h_{\mu}^{x}\left(\nabla_{\mathbf{y|x}^t}, \mathbf{x}^t\right) = \gamma_{t}\left(\nabla_{\mathbf{y|x}^t} + \nabla\log \text{p}_{\theta}(\mathbf{x}^t)\right)
        \end{equation*}

    
        Inspired by RIM, we propose the network shown in Fig.~\ref{fig:version2}. It uses the likelihood model $\text{p}(\mathbf{y|x}),$ updated at every iteration, as input to the Encoder-Predictor block which updates the state and symbol estimates as follows:
        \begin{equation*}\label{eq:9} \tag{9}
            S^{t+1} = S^{t} + h_{\theta_e^{t+1}}^s(S^{t}, \hat{X}^t, \nabla_{\mathbf{y}|\hat{\mathbf{x}}^t})
        \end{equation*}
        \begin{equation*}\label{eq:10} \tag{10}
            \hat{\mathbf{x}}^{t+1}_{i} = h_{\theta_p^{t+1}}^x\left(\mathbf{s}^{t+1}_{i}, \hat{\mathbf{x}}^{t}_{i}, \nabla_{\mathbf{y}|\hat{\mathbf{x}}^t}^{i}\right), \text{                } \left[\forall i \in N_{tr}\right]
        \end{equation*}
        where $S^t=\{\mathbf{s}^t_i\}_{i\in N_{tr}}$, $\hat{X}^t=\{\hat{\mathbf{x}}^t_i\}_{i\in N_{tr}}$, with $\mathbf{s}^t_i,\mathbf{x}^t_i$ denoting the state and symbol estimate for user $i$ at iteration $t$. 
        The $h_{\theta_e^{t+1}}^s$ and $h_{\theta_p^{t+1}}^x$ are the update models for the $(t+1)^{th}$ encoder and decoder module respectively with $\theta_e^{t+1}$ and $\theta_p^{t+1}$ as their trainable parameters. RE-MIMO builds its current prediction on the previous reconstruction, a hidden memory state, and the gradient of the log-likelihood term which injects information about the known generative process and measures how well the network is currently able to reproduce the measurements.
        
        To guide the model's predictions at each iteration in the right direction, we define the total loss function as a weighted mean of prediction losses at each iteration step. Also, to make the magnitude of the total loss independent of the number of transmitter antennas involved, we divide the loss function by $N_{tr}$. The final loss is given by
        \begin{equation*}\label{eq:11} \tag{11}
            \mathcal{L}^{final}(\theta) = \frac{1}{M} \sum_{m=1}^M \frac{1}{N_{tr}^{m}}\sum_{t=1}^{T} \omega_t \sum_{i=1}^{N_{tr}^{m}}  \mathcal{L}(\hat{\mathbf{x}}^{m,t}_i(\theta), x_i^m)
        \end{equation*}
        with $\sum_t \omega_t = 1$. Here, $m$ indexes different training samples, $N_{tr}^{m}$ is the number of transmitters in training sample $m$, and $\mathcal{L}(\hat{\mathbf{x}}^{m,t}_i(\theta), x_i^m)=CE(\hat{\mathbf{x}}^{m,t}_i(\theta), \text{one\_hot}(x_i^m))$ refers to cross entropy loss between the predicted probability distribution and the one-hot encoded true label.

    \subsection{The Likelihood module}
        The likelihood module captures the model information and any prior knowledge about the measurement process. The log-likelihood gradient of the forward MIMO model is given by:
        \begin{equation*}\label{eq:12} \tag{12}
            \nabla \log \text{p}(\mathbf{y}|\mathbf{x}, \mathbf{H}) \equiv \nabla_{\mathbf{y}|\mathbf{x}} =  \frac{\mathbf{H}^{H}(\mathbf{y}-\mathbf{H} \mathbf{x})}{\sigma^2} \equiv \frac{\mathbf{H}^{H} \cdot \delta \mathbf{y}}{\sigma^2}
        \end{equation*}
        where we have used $\nabla_{\mathbf{y}|\mathbf{x}}$ to denote $\nabla \log \text{p}(\mathbf{y}|\mathbf{x}, \mathbf{H})$ and $\delta \mathbf{y}$ as the residual vector $\mathbf{y}-\mathbf{H} \mathbf{x}$. The likelihood model is also permutation equivariant. Denoting $\mathcal{P}$ as the permutation matrix, this can be derived as follows:
        \begin{equation*}\label{eq:13} \tag{13}
            \nabla_{\mathbf{y}|\mathbf{x}}(\mathbf{H}',\mathbf{x}') \equiv \nabla_{\mathbf{y}|\mathbf{x}}(\mathbf{H}\mathcal{P}, \mathcal{P}^{T}\mathbf{x}) = \frac{ (\mathbf{H\mathcal{P}})^{H}(\mathbf{y}-\mathbf{H}\mathcal{P}\mathcal{P}^{T} \mathbf{x})}{\sigma^2}
        \end{equation*}
        \begin{equation*}\label{eq:14} \tag{14}
            \nabla_{\mathbf{y}|\mathbf{x}}(\mathbf{H}',\mathbf{x}') = \frac{\mathcal{P}^T\mathbf{H}^{H}(\mathbf{y}-\mathbf{H} \mathbf{x})}{\sigma^2} \equiv \mathcal{P}^T\nabla_{\mathbf{y}|\mathbf{x}}(\mathbf{H},\mathbf{x})
        \end{equation*}
        
        In RE-MIMO, each user gets only its component of log-likelihood gradient which is given by
        \begin{equation*}\label{eq:15} \tag{15}
            \frac{\partial \log \text{p}(\mathbf{y}|\mathbf{x,H})}{\partial x_i} \equiv \nabla_{\mathbf{y}|\mathbf{x}}^{i} = \frac{\mathbf{h}_{i}^{H} \cdot \delta \mathbf{y}}{\sigma^2}
        \end{equation*}
        where $\nabla_{\mathbf{y}|\mathbf{x}}^{i}$ denotes the $i^{th}$ component of log-likelihood gradient vector $\nabla_{\mathbf{y}|\mathbf{x}}$. We inject the gradient information into the detector in terms of its constituent components, $\mathbf{h}_{i}, \delta \mathbf{y},$ and $\sigma$. The first is permutation equivariant, and the latter two are permutation invariant. We leave it to the network's discretion in deciding how to effectively combine/utilize the information.
        
        The magnitude of the log-likelihood gradient components scales with the number of users in the MIMO system. So, we normalize the components individually before passing it to the modules, in such a way that the variance of the components remains constant, i.e., $\delta \mathbf{y} \rightarrow \frac{\delta \mathbf{y}}{\sqrt{2N_{tr}}}$ and $\sigma \rightarrow \frac{\sigma}{\sqrt{2N_{tr}}}$. A factor of two appears in the denominator because the complex MIMO system is translated to its equivalent real valued representation, due to which every variable has two components, real and imaginary.

        \begin{figure}[!t]
            \centering
            \includegraphics[width=3.4in]{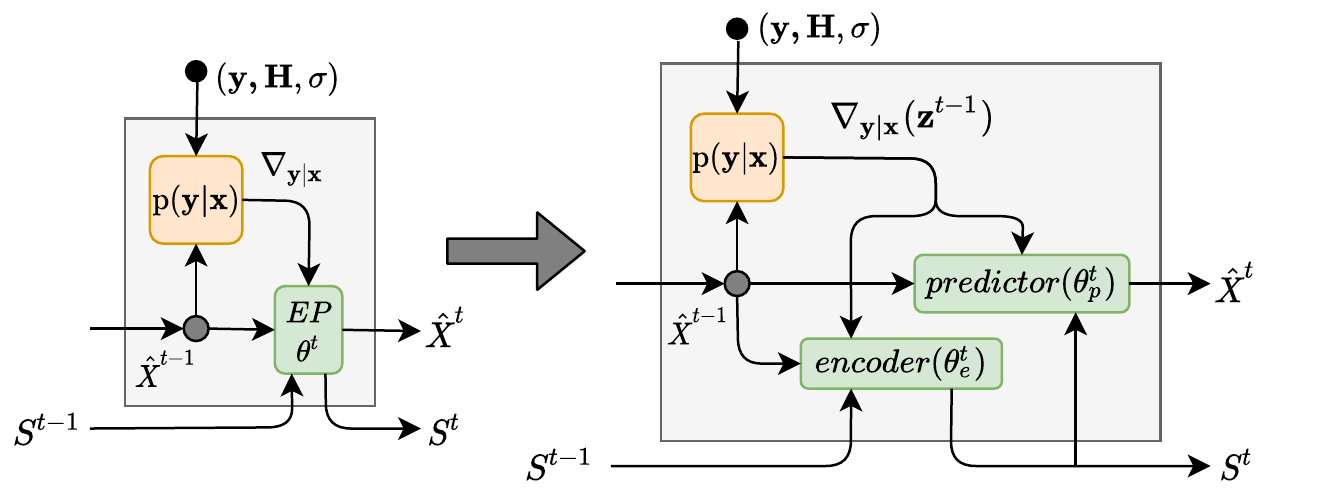}
            \caption{\textit{Block description of one iteration step.} The schematic on the right is a detailed illustration of the schematic on the left. $\theta_{e}^{t}$ and $\theta_{p}^{t}$ denote weights associated with $t^{th}$ encoder and predictor modules respectively. The encoder module jointly updates the state variables, based on which the predictor module produces a prediction individually for each user.}
            \label{fig:EP}
        \end{figure}
        The log-likelihood gradient requires the computation of the residual vector $\delta \mathbf{y}$, which in turn requires an intermediate signal reconstruction. Although the proposed framework reconstructs the transmitted signal in terms of a probability distribution over symbols in $\mathcal{X}$, we can obtain a numerical estimate of the intermediate signal reconstruction $\mathbf{z}^t$ at time step $t$ from the output of the predictor module as follows:
        \begin{equation*}\label{eq:16} \tag{16}
            z^{t}_{i} = \sum_{\mathcal{X}_{j} \in \mathcal{X}} p(x_i=\mathcal{X}_j|\textbf{y},\mathbf{H},\hat{\mathbf{x}}^{t}_{i})\cdot \mathcal{X}_{j} = \sum_{\mathcal{X}_{j} \in \mathcal{X}} p_{ij}^{t} \cdot \mathcal{X}_{j}
        \end{equation*}
        which is a probability weighted sum of the symbols in $\mathcal{X}$. Unlike $x_i$, $z^t_{i}$ is not constrained to lie in the constellation set $\mathcal{X}$ and is used to compute the residual vector, $\delta \mathbf{y}^{t} = \mathbf{y}-\mathbf{H} \mathbf{z}^{t}$. The likelihood model $\text{p}(\mathbf{y|x})$ computes the log-likelihood gradient $\nabla \log \text{p}(\mathbf{y}|\mathbf{z}^{t}, \mathbf{H})$ based on the previous prediction as depicted in Fig.~\ref{fig:EP}.
  
    \subsection{The Encoder module}
        The encoder module is an important component of our RE-MIMO framework which deals with user interaction in a permutation equivariant manner. We first discuss the factors leading to the choice of the network, a few salient features of the network, and specifics about the receiver design.
        \subsubsection{Choice of Network structure} 
            The encoder module is a  transducer that transforms a set of input state variables $\left\{\mathbf{s}^{t}_{i}\right\}_{i \in N_{tr}}$ into an output set of updated state variables $\left\{\mathbf{s}^{t+1}_{i}\right\}_{i \in N_{tr}}$. From the network design point of view, it is useful to view the indices assigned to the individual users as sequence indices. This enables one to benefit from the recent developments in seq-to-seq transduction tasks in traditional machine learning fields like speech recognition, neural machine translation, etc. In the communication task, the user index assignment is arbitrary and so the network must account for multi-user interaction independent of the sequence order. The network needs to capture interference between users that are located far apart in the sequence $S^t$. The encoder must also deal with a varying number of users, i.e., variable sequence length, and in a permutation equivariant manner.
            
            For seq-to-seq transduction tasks, neural network (NN) variants such as convolutional neural networks (CNNs), RNNs, and the transformer networks are often deployed. As RNNs are intended for ordered sequential data, their outputs are highly sensitive to the order of the input sequence. Also, their ability to capture dependencies between two entities depends on the path length between them which makes it difficult for RNNs to model long distance dependencies. CNNs are capable of modeling correlations but only in a fixed neighborhood (the receptive field), the size of which depends on the kernel size $k$ and the number of layers. A single convolutional layer with kernel size $k<N_{tr}$ does not connect all pairs of input and output positions as required for our encoder module. Also, both the CNNs and RNNs do not satisfy the permutation equivariance property. Self-attention networks on the other hand, such as transformers, connect all pairs of input and output values and are permutation equivariant. Unlike RNNs, they allow parallelization for most of the computations involved which makes training efficient and also more viable for real-time detection requirements during inference. We refer our readers to~\cite{transformers} for an in-depth discussion. More importantly, while decoding the symbol transmitted by a user, the detector must figure out which of the other co-transmitters it should pay attention to, and successively account for their interference. Classical detection algorithms achieve this by using parallel/successive interference cancellation to reduce interference~\cite{pic_1, pic_2}. Based on the above mentioned considerations, transformers are a natural choice.
        
        \subsubsection{Transformer Network}
            \begin{figure*}[!t]
                \centering                \includegraphics[height=80mm]{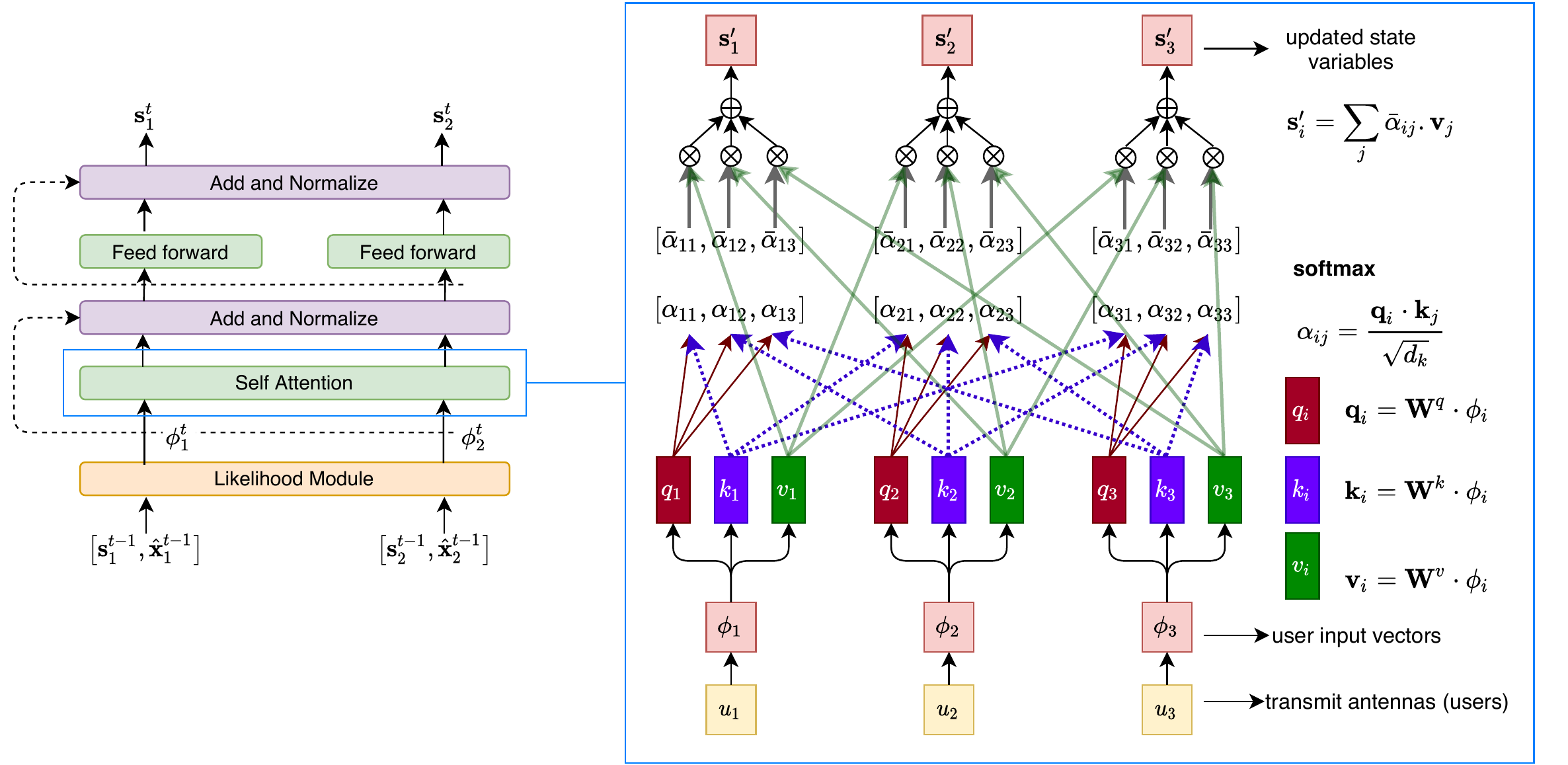}
                \caption{\textit{The Encoder Module}. The Encoder module jointly updates the state variables of all the users based on the previous state variables, preceding reconstruction, and the log-likelihood gradient components. The flow diagram on the right zooms into the self-attention layer of the encoder module and demonstrates the self-attention mechanism of the Transformer network involving three transmitters. The self-attention mechanism can be extended to an arbitrary number of transmitters.}
                \label{fig:transformer}
            \end{figure*}
            The transformer network~\cite{transformers}, shown in Fig.~\ref{fig:transformer}, is used in our encoder module to implement the self-attention function.
            Self-attention, also known as intra-attention, is an attention mechanism relating different positions of a single sequence in order to compute a representation of the same sequence. An attention function can be described as mapping a query and a set of key-value pairs to an output, where the query, keys, values, and output are all vectors. The output is computed as a weighted sum of the values, where the weight assigned to each value is computed by a compatibility function of the query with the corresponding key.
            
            The Transformer uses scaled dot-product self-attention, where we train three sets of matrices called query, key, and value matrices, denoted by $\mathbf{W}^q, \mathbf{W}^k, \mathbf{W}^v$ respectively. For every transmitter (user) $u_i$, we have defined an input vector $\phi_i$. We calculate query, key, and value vectors for each of the users as follows: $\mathbf{q}_i=\mathbf{W}^{q} \cdot \phi_i$, $\mathbf{k}_i=\mathbf{W}^{k} \cdot \phi_i$, and $\mathbf{v}_i=\mathbf{W}^{v} \cdot \phi_i$.
            
            $\mathbf{W}^{q}, \mathbf{W}^{k}$, and $\mathbf{W}^{v}$ are the trainable parameters of the self-attention layer. Every self-attention layer has its own set of the query, key, and value matrices. Attention scores $\mathbf{\alpha}$ are defined as an inner product of the query vector with the key vector divided by the squared root of the dimensionality of the query vector. Every user generates its attention towards all co-transmitters by computing the dot product of its query vector with the key vector of the target user. This gives us a matrix $\mathbf{A}$ of attention scores, where $\alpha_{ij}$ shows the attention given by user i to user j. We apply a softmax function on the attention matrix $\mathbf{A}$ row-wise, such that the attention scores sum to one for every user. The softmax score $\bar{\alpha}_{ij}$ determines how much the user $j$ contributes towards computing the updated state variable for user $i$. The updated state variable $\mathbf{s}'_i$ is computed by summing the value vectors weighted by their corresponding attention score.
            
            Multi-head attention introduced in~\cite{transformers}, is an extension of self-attention where instead of calculating just one attention matrix $\mathbf{A}$, multiple attention matrices $\mathbf{A}_i^{s}$ are created in parallel. Similar to CNNs, where one convolutional layer has multiple convolutional kernels, one multi-head self-attention unit has multiple self-attention sub-units in parallel. Each individual self-attention sub-unit has its own set of trainable matrices given by $\mathbf{W}^{q}_{i}, \mathbf{W}^{k}_{i}$, and $\mathbf{W}^{v}_{i}$. The outputs of the individual self-attention sub-units are concatenated together and linearly transformed into the required dimension by a learned projection matrix $\mathbf{W}^{o}$.
            
            A Multi-head attention module with $n$ heads allows each head to attend to a different subspace. With multiple heads, the self-attention layer generates multiple outputs which are then concatenated and linearly transformed into the required dimension to get the final output.
            \begin{equation*}\label{eq:17} \tag{17}
                \text{MultiHead self-attention}(\Phi) = \text{Concat}(\text{head}_1,..., \text{head}_n)\mathbf{W}^o
            \end{equation*}
            \begin{equation*}\label{eq:18} \tag{18}
                \text{where head}_i = \text{self-attention}(\mathbf{Q}_{i}, \mathbf{K}_{i}, \mathbf{V}_{i})
            \end{equation*}
            
            Here, $\Phi, \mathbf{Q}_{i}, \mathbf{K}_{i}, \text{ and } \mathbf{V}_i$ represent the set of individual user input, queries, keys, and values vectors stacked together into matrix form, respectively and is given by $\Phi = \{\phi_{j}\}_{j \in N_{tr}}, \text{ }\mathbf{Q}_{i} = \Phi \mathbf{W}^{q}_{i}, \text{ }\mathbf{K}_{i} = \Phi \mathbf{W}^{k}_{i}, \text{ and } \mathbf{V}_i = \Phi \mathbf{W}^{v}_{i}$. The self-attention sub-unit can be depicted as:
            \begin{equation*}\label{eq:19} \tag{19}
                \text{self-attention}(\mathbf{Q}_{i}, \mathbf{K}_{i}, \mathbf{V}_i) = \text{softmax}(\frac{\mathbf{Q}_{i} \mathbf{K}_{i}^{T}}{\sqrt{d_k}}) \mathbf{V}_{i}
            \end{equation*}
            
        \subsubsection{Module details}
            Our encoder module, illustrated in Fig.~\ref{fig:transformer}, is an adaptation of the encoder module from the transformer network~\cite{transformers}. It is composed of two sub-layers, the first is a multi-head self-attention layer, and the second is a position-wise fully connected feed-forward network. There is a residual connection~\cite{resnet} around each of the two sub-layers, followed by layer normalization~\cite{layernorm}. The output of each sub-layer is $\text{LayerNorm}(\mathbf{s}^{t-1}_i+\text{Sublayer}(\mathbf{\phi}^t_i))$, where $\text{Sublayer}(\mathbf{\phi}^t_i)$ is the function implemented by the sub-layer itself. The second sub-layer in the encoder module is a fully connected point-wise feed-forward neural network with one hidden layer of size $4d_s$ and the ReLU activation function.
       
            Our encoder module deploys a slightly modified transformer to learn an iterative inference algorithm. The individual user input to the $t^{th}$ encoder module, $\mathbf{\phi}^t_i$ is a concatenation of the state variable $\mathbf{s}$, previous prediction for $\mathbf{x}$, and normalized components of the log-likelihood gradient, $\mathbf{h}$ and $\delta\mathbf{y}$. It is given by
            \begin{equation*}\label{eq:20} \tag{20}
                \mathbf{\phi}^t_i = \left[\mathbf{s}^{t-1}_i, \hat{\mathbf{x}}^{t-1}_{i}, \frac{\delta \mathbf{y}^{t-1}}{\sqrt{2N_{tr}}}, \mathbf{h}_{i}\right]
            \end{equation*}
            As $\sigma$ is the same for every user, it does not participate in modeling interactions among users and hence, it is not included in the encoder input vector.
            
            As the encoder module jointly updates all state variables of every user, it takes the set $\{\phi^t_i\}_{i \in N_{tr}}$ as input and outputs the updated state variable set $S^t={\{\mathbf{s}^t_i}\}_{i \in N_{tr}}$. 
            Unlike the encoder in the Transformers Network, our encoder module does not have the same input and output dimensionality. To account for this, the dimensionality of the projection and parameter matrices are adjusted such that $\mathbf{W}^q_i \in \mathbb{R}^{d_\phi \times d_k}, \mathbf{W}^k_i \in \mathbb{R}^{d_\phi \times d_k}, \mathbf{W}^v_i \in \mathbb{R}^{d_\phi \times d_v}$, and $\mathbf{W}^o \in \mathbb{R}^{nd_v \times d_s}$. The $d_{\phi}, d_s, d_k, \text{ and }d_v$ represent the dimensionality of input vectors, state variables, key, and value vectors respectively. In our implementation, we use $n=8$ parallel attention heads in the multi-head self-attention layer and for each attention head $d_k, d_v=d_{\phi}/n$. The dimension of the state variable $d_s$ is chosen to be $512$ in our implementation. The input dimension $d_\phi$ is automatically determined by the dimensions of its constituent components as mentioned in (\ref{eq:20}) which in turn is determined by the dimension of the state variable $d_s$, the modulation scheme used, and the number of receivers $N_r$.

    \subsection{The Predictor module}
        \begin{figure}[!t]
            \centering
            \includegraphics[height=45mm]{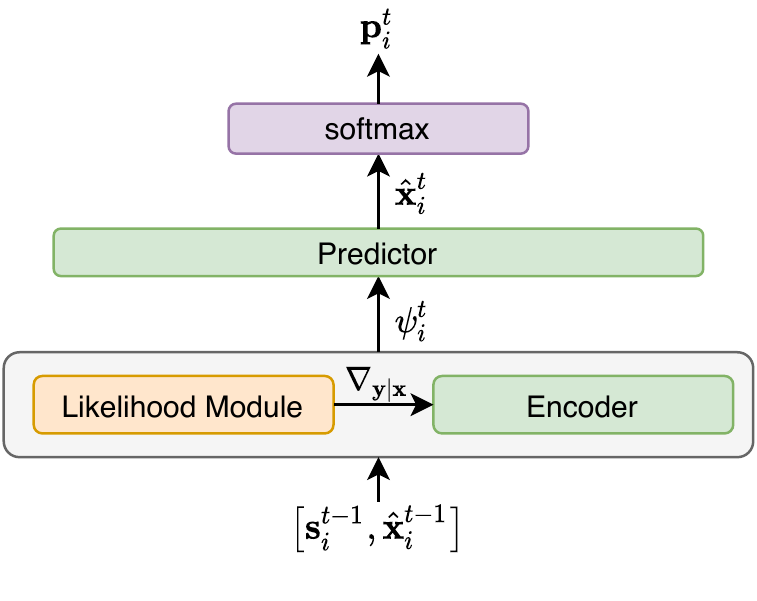}
            \caption{\textit{The Predictor module}. The predictor module outputs the updated signal reconstruction for each user individually, based on its previous state variable, preceding reconstruction, and the log-likelihood gradient components. The likelihood and encoder modules are depicted just to illustrate the data flow through modules.}
            \label{fig:predictor}
        \end{figure}
        The predictor module, shown in Fig.~\ref{fig:predictor}, is a classifier that outputs a probability distribution over a constellation set $\mathcal{X}$ individually for every transmitted symbol. The pre-softmax output of the predictor module for user $i$ at time $t$ is given by $\hat{\mathbf{x}}^t_i$ which is then forwarded through a softmax operation to get the final output vector $\mathbf{p}^t_i$, i.e., $\mathbf{p}^t_i=softmax(\hat{\mathbf{x}}^t_{i})$. Similar to the input to the encoder, the individual user's input to the $t^{th}$ predictor block, $\psi_i^t$ is a concatenation of the state variable $\mathbf{s}$, the pre-softmax prediction variable $\mathbf{x}$ from the preceding block, and normalized components of the log-likelihood gradient. It is given by 
        \begin{equation*}\label{eq:21} \tag{21}
            \mathbf{\psi}^t_i = \left[\mathbf{s}^{t}_i, \hat{\mathbf{x}}^{t-1}_{i}, \frac{\delta \mathbf{y}^{t-1}}{\sqrt{2N_{tr}}}, \mathbf{h}_{i}, \frac{\sigma}{\sqrt{2N_{tr}}}\right]
        \end{equation*}
        Here, we include $\sigma$ in the input vector as we want the predictor to accordingly adjust the variance of its predicted probability distribution. The predictor module is a fully connected feed-forward network with two hidden layers and ReLU activation. The same predictor network is applied independently for every transmitter. As can be seen from the input to the predictor module there is no explicit exchange of information between the transmitters in the predictor module. The parameters of the network are the same for all transmitters making the overall network permutation equivariant. Similar to the input of the encoder module, the input dimension $d_\psi$ is automatically determined by the dimensions of its constituent components as mentioned in (\ref{eq:21}). In our implementation, the dimensions of the first and second hidden layers of the predictor module are kept at $\frac{d_\psi}{2}$ and $\frac{d_\psi}{4}$ respectively.
        
    \subsection{Network Initialization}
        In this section, we present the \textit{Embedding Network} ($\gamma$) which is used to initialize the state variables for each transmitter. The embedding network is a point-wise feed-forward fully connected network with one hidden layer and ReLU activation. The input to $\gamma$ is a hand designed feature vector for every transmitter given by
        \begin{equation*}\label{eq:22} \tag{22}
            \mathbf{f}_i = \left[ \frac{\textbf{y}}{\sqrt{2N_{tr}}},  \mathbf{h}_i, \frac{\sigma}{\sqrt{2N_{tr}}}, TE\right]
        \end{equation*}
        Here, $\mathbf{f}_i$ represents the feature vector for the $i^{th}$ transmitter and $TE$ represents the \textit{Transmitters Encoding} vector described in sec.~\ref{sec:TE}. As the number of transmitters changes, the variance of $\mathbf{y}$ varies. To keep the variance at the input of the embedding network $\gamma$ constant, we normalize $\mathbf{y}$ by dividing it by $\sqrt{2N_{tr}}$. Since the RE-MIMO architecture is independent of the number of transmitters involved, to inject information about the number of transmitters, we include $TE$, as explained in sec.~\ref{sec:TE}, in our feature vector. The feature vectors are then forwarded through the embedding network $\gamma$ to get the initial state vector (embedding) for each transmitter. The output embedding from $\gamma$ is multiplied by $\sqrt{d_s}$ before forwarding it to the first EP block, i.e. $\mathbf{s}^0_{i} = \gamma(\mathbf{f}_i) \cdot \sqrt{d_s}$. The initial prediction vector $\hat{\mathbf{x}}_i^0$ is initialized as a zero vector. Only the first encoder-predictor block receives input from the $\gamma$ network, all other EP blocks take the output of the preceding block as an input. The dimensionality of the hidden layer is chosen to be $4d_s$. In our implementation, we keep the value of $d_s$ at $512$.
        
    \subsection{TE: Transmitters Encoding}
        \label{sec:TE}
        Positional encoding is used in transformer networks to identify the relative position of an entry in a sequence. Here, we use the positional encoding feature,
        referred to as \textit{Transmitters Encoding} ($TE$), in a novel way. $TE$
        encodes the number of transmitters involved in generating the received signal vector $\mathbf{y}$. To make every transmitter aware of the number of co-transmitters involved, $TE$ injects this information into the feature vector $\mathbf{f}_i$ of each of the transmitters. As the number of transmitters varies, $TE$ is intended to aid the network in terms of scaling its operations accordingly, such as scaling the log-likelihood gradient, the amount of interference, etc. The $TE$ vector is taken from the positional encoding proposed in~\cite{transformers} and is adjusted for our use case. The $TE$ vector is given by:
        \begin{equation*}\label{eq:23} \tag{23}
            TE_{(N_{tr},2i)} =\sin{\left(N_{tr}/(2N_r)^{2i/d_{TE}}\right)}/ \sqrt{d_s}
        \end{equation*}
        \begin{equation*}\label{eq:24} \tag{24}
            TE_{(N_{tr},2i+1)} =\cos{\left(N_{tr}/(2N_r)^{2i/d_{TE}}\right)}/ \sqrt{d_s}
        \end{equation*}
        Here, $i$ denotes the $i^{th}$ component (dimension) of the $TE$ vector, and $d_{TE}$ represents the dimensionality of $TE$, which in our implementation is kept equal to the number of receivers, i.e., $d_{TE}=N_r$. Each dimension of $TE$ corresponds to a sinusoid. The wavelengths form a geometric progression from $2\pi$ to $2NR\cdot2\pi$. The rationale behind using sinusoidal embeddings instead of learned embeddings or one-hot encoding is that it may assist the network to interpolate or extrapolate to MIMO systems with the number of transmitters that have never been encountered during training. It should be noted that $TE$ is the same for every user.
        
\section{Experiments}
\label{sec:experiments}
    In this section, we present numerical results to evaluate and compare the performance of RE-MIMO with other well established MIMO detection schemes. We use the symbol error rate (SER) as the performance evaluation metric in our experiments. We perform a separate experiment to understand the inner working dynamics of RE-MIMO. The detection schemes used in the experiment for comparison purposes are listed below. Then, we discuss the generation of data on which these schemes are trained and/or tested. Below we discuss the implementation details of the detection schemes used in our experiments.
    \begin{itemize}
        \item \textbf{MMSE}: Classical linear receiver that inverts the signal by applying the channel-noise regularized pseudoinverse of the channel matrix.
        \item \textbf{AMP}: AMP algorithm as given in~\cite{amp} and implemented with $50$ iterations. It was verified in~\cite{mmnet} that increasing the number of iterations does not lead to an increase in performance.
        \item \textbf{SDR}: Semidefinite programming with rank $1$ relaxation and implemented using an efficient interior point method~\cite{sdr_results}.
        \item \textbf{V-BLAST}: Multi-stage successive interference cancellation BLAST algorithm and using ZF detector at the detection stage as presented in~\cite{pic_3}.
        \item \textbf{ML}: The optimal ML solver for (\ref{eq:2}) implemented using Gurobi~\cite{gurobi} optimization package.
        \item \textbf{DetNet}: Deep learning architecture with $3N_{tr}$ layers as introduced in \cite{deepmimo}. We experiment with complex MIMO systems, hence, we double the dimensionality of latent variables ($\mathbf{z}, \mathbf{v}$) to accommodate the increased complexity. A separate DetNet is trained for each combination of $\{N_{r}, N_{tr}, \text{QAM-order}\}$.
        \item \textbf{OAMPNet}: The OAMP based iterations unrolled in $10$ layers as proposed in~\cite{oampnet}. Each layer requires computing a matrix pseudoinverse and has $2$ learnable parameters. Like DetNet, a separate OAMPNet is trained for each combination of $\{N_{r}, N_{tr}, \text{QAM-order}\}$.
        \item \textbf{OAMPNet-2}: The OAMP based iterations unrolled in $10$ layers as proposed in~\cite{oamp_net_2}. Each layer requires computing a matrix pseudoinverse and has $4$ learnable parameters. Similar to OAMPNet, a separate OAMPNet-2 is trained for each combination of $\{N_{r}, N_{tr}, \text{QAM-order}\}$.
        \item \textbf{RE-MIMO}: Recurrent permutation equivariant neural detector as explained in sec.~\ref{sec:RE-MIMO}. Unlike DetNet and OAMPNet-2, we train a single detector for a given combination of $\{N_{r}, \text{QAM-order}\}$, i.e., a single RE-MIMO handles all the values of $N_{tr}$ in a pre-defined range.
    \end{itemize}
    The implementation code for AMP, SDR, V-BLAST, and ML is taken from the GitHub repository of \cite{mmnet}. Our repository containing PyTorch implementation of learning-based schemes and the experiments presented in the paper is available at \url{https://github.com/krpratik/RE-MIMO}.
    
    In our experiments, we consider massive MIMO systems where the number of transmitters is smaller than the number of receivers at the BS. Unless mentioned otherwise, our experiments consider MIMO configurations with $N_{r}=64$ and system size ratios $(\frac{N_{tr}}{N_{r}})$ belonging to the range, $\frac{N_{tr}}{N_r} \in \left[\frac{1}{4}, \frac{1}{2}\right]$. For system size ratios below the aforementioned range, the performance of traditional detection schemes (such as V-BLAST) is at par with the optimal detection schemes and can effectively be deployed for symbol detection purposes.
    
    The performance of detection algorithms highly depends on the type of MIMO channel. Therefore, we evaluate the performance of RE-MIMO under both i.i.d. Gaussian and correlated Rayleigh fading channels with perfect CSI. The channel noise $\mathbf{n}$ follows a zero-mean i.i.d. Gaussian distribution whose variance is related to the SNR as per the formula
    \begin{equation*}\label{eq:25} \tag{25}
        \text{SNR} = \frac{\mathbb{E}\left[||\mathbf{Hx}||_{2}^{2}\right]}{\mathbb{E}\left[||\mathbf{n}||_{2}^{2}\right]}
    \end{equation*}
    To generalize the scope of our findings, we experiment with two different orders of QAM modulation scheme, QAM-$16$ and QAM-$64$.
    \subsection{Rayleigh MIMO channel}
    \label{sec:i.i.d}
        In this subsection, we study i.i.d. Gaussian channels where each element of $\mathbf{H}$ is sampled from $h_{ij} \sim  \mathcal{C} \mathcal{N}\left(0, 1/N_r\right)$, i.e., each element of $\mathbf{H}$ follows a zero-mean circularly-symmetric Gaussian distribution with variance $1/N_r$.

        \begin{figure*}[!t]
            \centering
            \subfloat[Case I]{\includegraphics[width=3.2in]{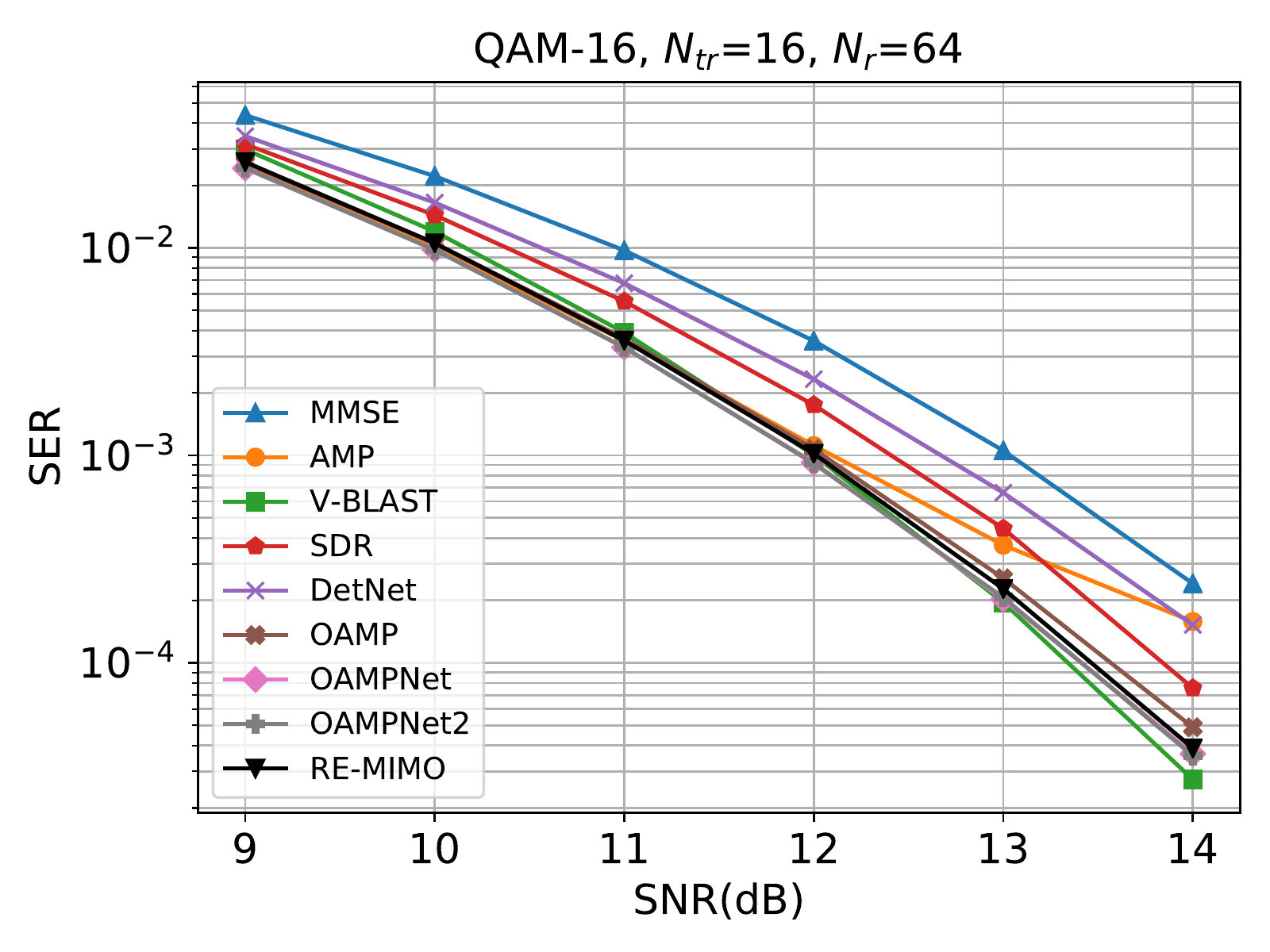}%
            \label{fig:qam_16_NT_16}}
            \hfil
            \subfloat[Case II]{\includegraphics[width=3.2in]{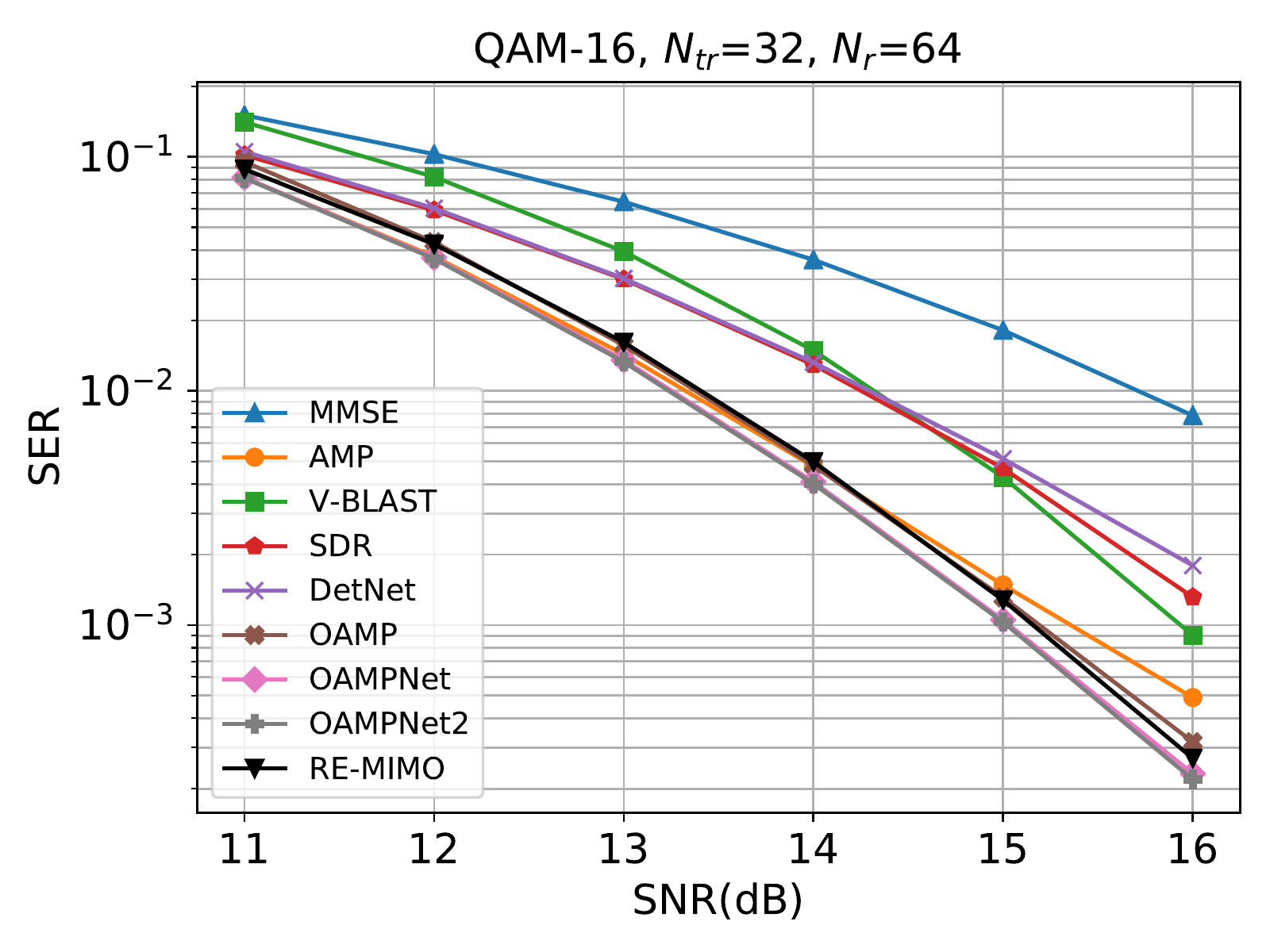}%
            \label{fig:qam_16_NT_32}}
            \caption{SER vs. SNR for $16$, $32$ transmit antennas and $64$ receive antennas on i.i.d. Gaussian channels with QAM-16 modulation.}
            \label{fig:qam_16}
        \end{figure*}
        
        \begin{figure*}[!t]
            \centering
            \subfloat[Case I]{\includegraphics[width=3.2in]{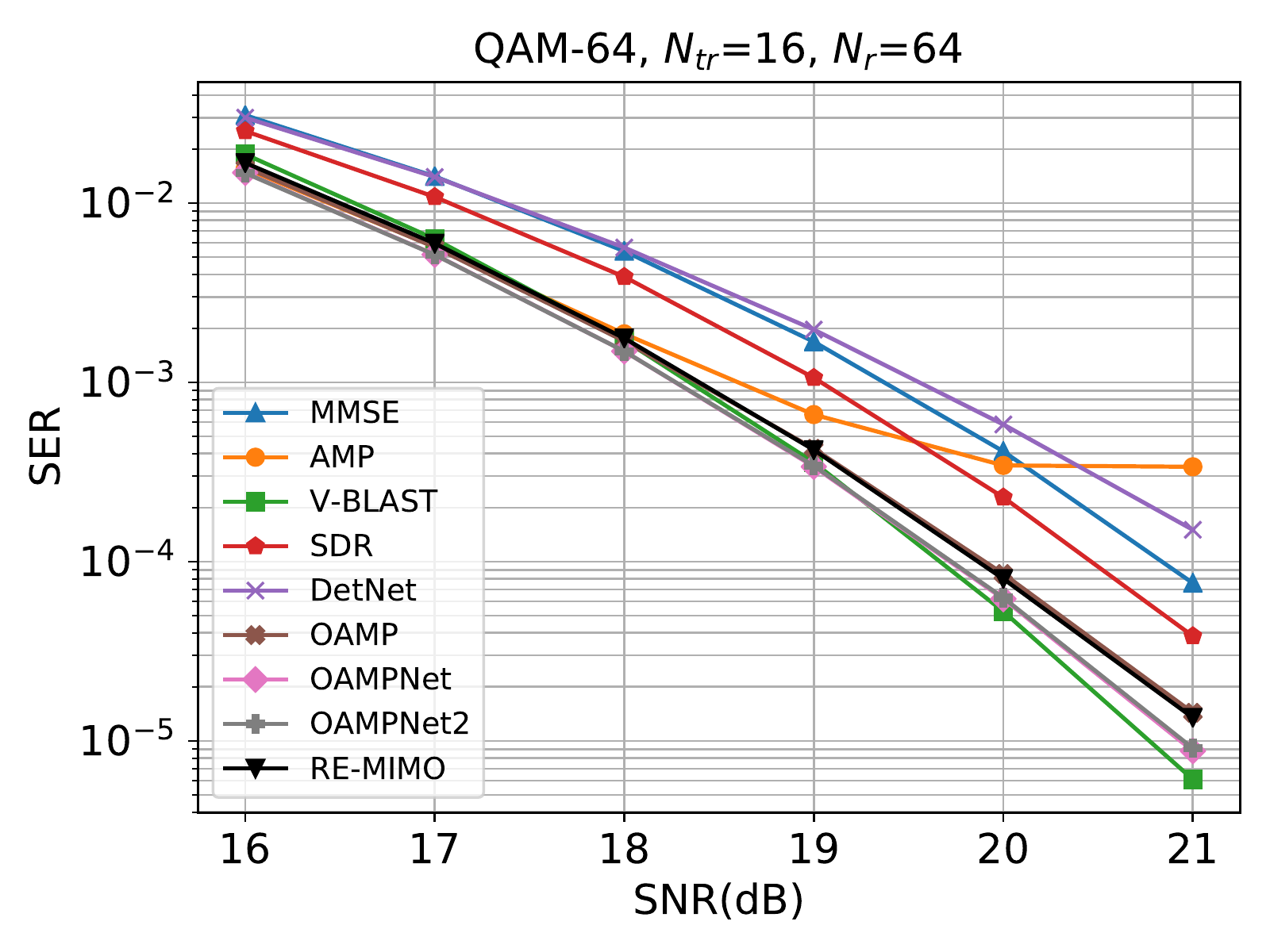}%
            \label{fig:qam_64_NT_16}}
            \hfil
            \subfloat[Case II]{\includegraphics[width=3.2in]{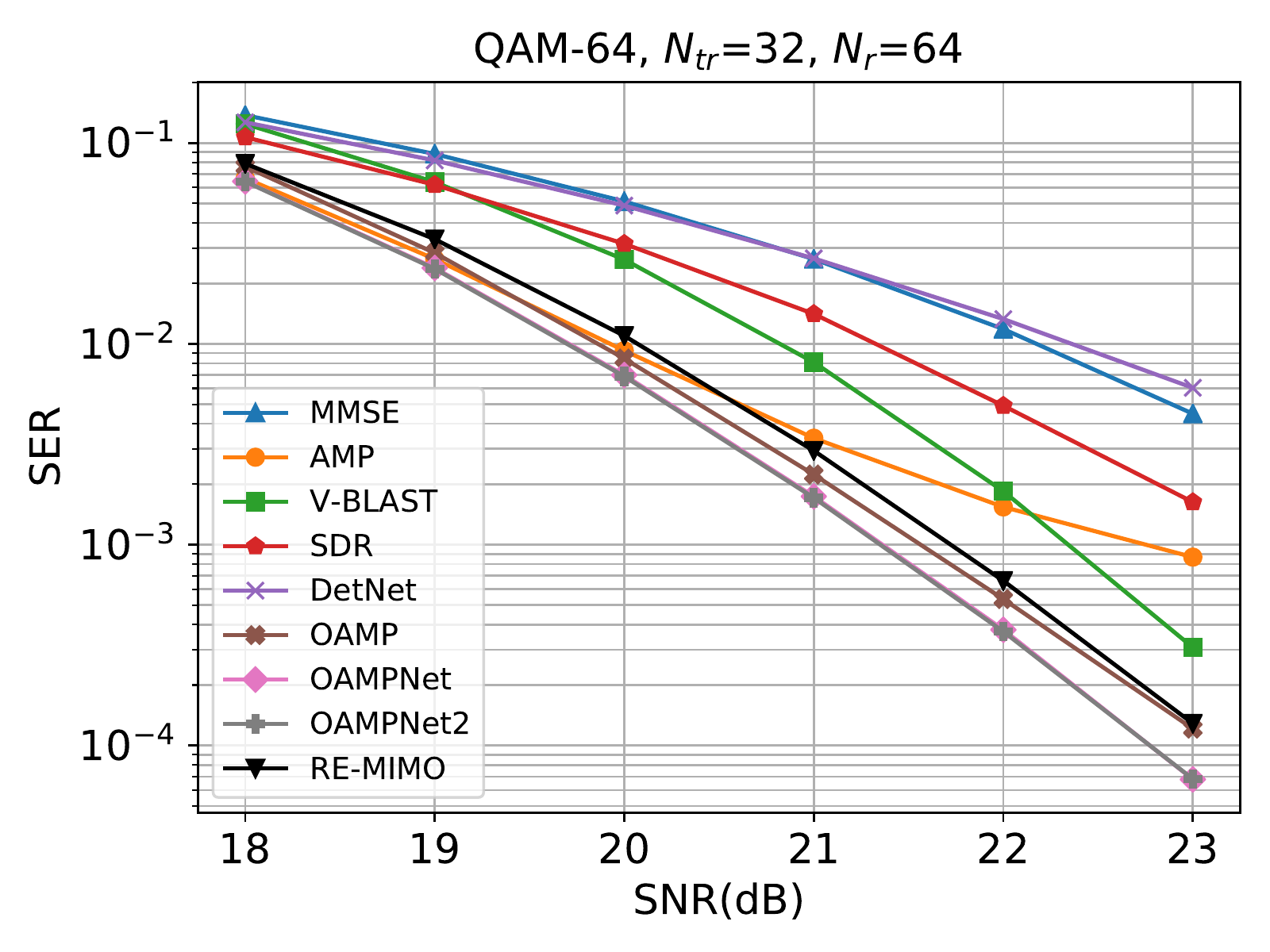}%
            \label{fig:qam_64_NT_32}}
            \caption{SER vs. SNR for $16$, $32$ transmit antennas and $64$ receive antennas on i.i.d. Gaussian channels with QAM-64 modulation.}
            \label{fig:qam_64}
        \end{figure*}
        
        Fig.~\ref{fig:qam_16} compares the SER of RE-MIMO with other detection schemes for two system size ratios, $0.25 \text{ and } 0.50$, on the QAM-16 modulation scheme. Our single RE-MIMO, which is simultaneously trained for all the values of $N_{tr}$ in the range, performs on par with the OAMPNet-2 which is optimal for i.i.d. Gaussian channels and which requires an exclusive network and separate training for each value of $N_{tr}$. Similarly, Fig.~\ref{fig:qam_64} depicts the SER performance for the same two system size ratios as above but with QAM-64 modulation. A single RE-MIMO again performs almost at par with the optimal detectors for i.i.d. channels trained separately for $N_{tr}=16$ and $N_{tr}=32$. For both the QAM-16 and QAM-64 modulation orders, we train RE-MIMO with $12$ EP blocks.
        
        Although RE-MIMO gracefully handles all the values of $N_{tr}$ in the range, we illustrate the performance only at the two boundary values of the range because the other learning-based schemes require separate training and storage of parameters for networks exclusively meant for a fixed number of users.
        
    \subsection{Correlated MIMO channel}
        Here, we consider correlated MIMO channels modeled by the Kronecker model
        \begin{equation*}\label{eq:26} \tag{26}
            \mathbf{H} = \mathbf{R}_{R}^{1/2}\mathbf{H}_{w}\mathbf{R}_{T}^{1/2}
        \end{equation*}
        where $\mathbf{H}_{w}$ is the Rayleigh fading channel matrix, $\mathbf{R}_{R}$ and $\mathbf{R}_{T}$ are the spatial correlation matrix at the receiver and transmitter side respectively which is generated according to the exponential correlation model~\cite{correlated} with correlation coefficient $\rho$. We also consider the scenario where the transmitters are located in different places and can be considered independent of each other. In that case, we do not expect any correlation between the channels of the different transmitters and hence, do not post multiply $\mathbf{H}_w$ by $\mathbf{R}_{T}^{1/2}$ in (\ref{eq:26}). We refer to the latter as a partially correlated case and the former as a correlated case.
        
        \begin{figure*}[!t]
            \centering
            \subfloat[Case I]{\includegraphics[width=3.2in]{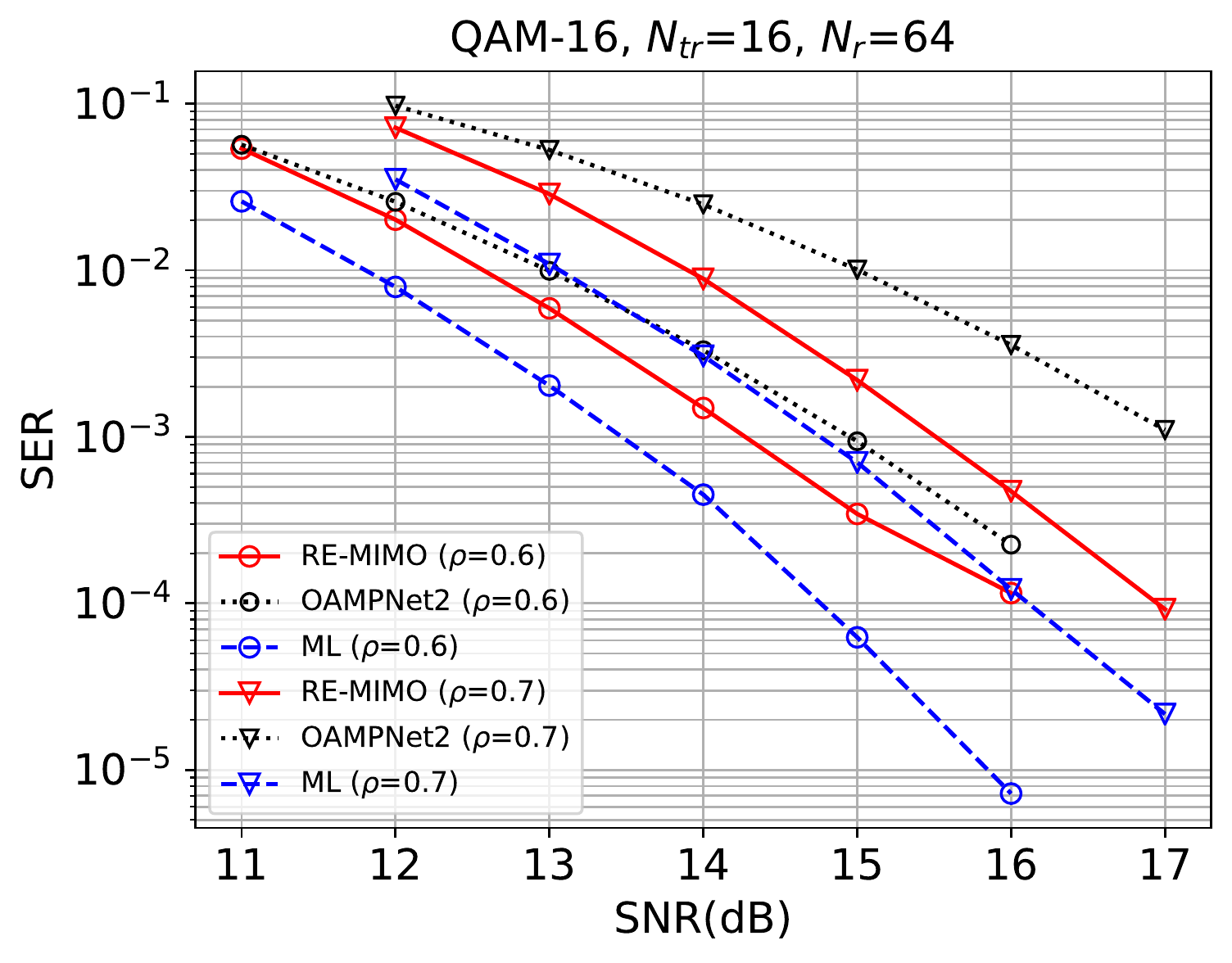}%
            }
            \hfil
            \subfloat[Case II]{\includegraphics[width=3.2in]{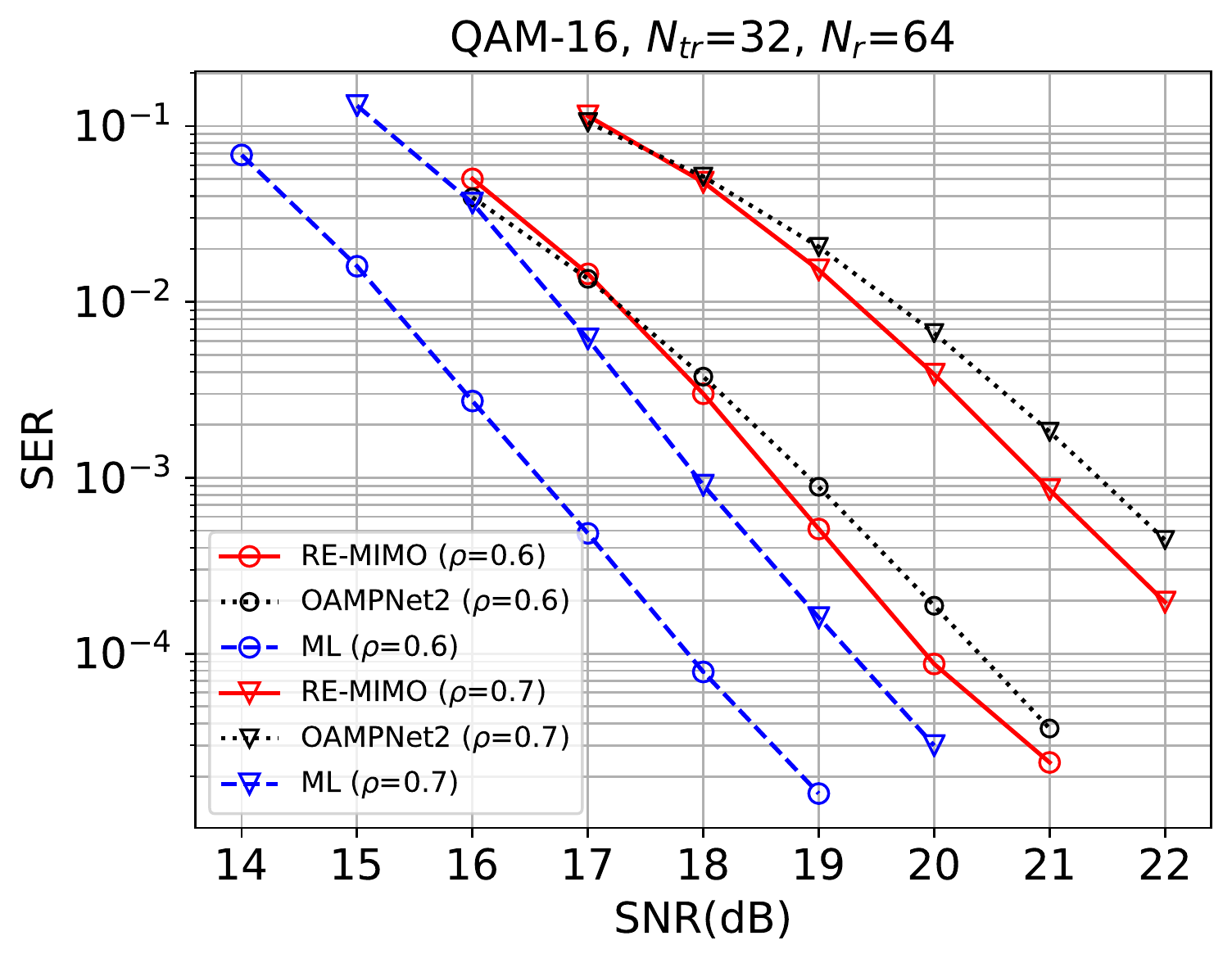}%
            }
            \caption{SER vs. SNR for $16$, $32$ transmit antennas and $64$ receive antennas on correlated Rayleigh channels channels with QAM-16 modulation.}
            \label{fig:full_corr_qam_16}
        \end{figure*}
        
        \begin{figure*}[!t]
            \centering
            \subfloat[Case I]{\includegraphics[width=3.2in]{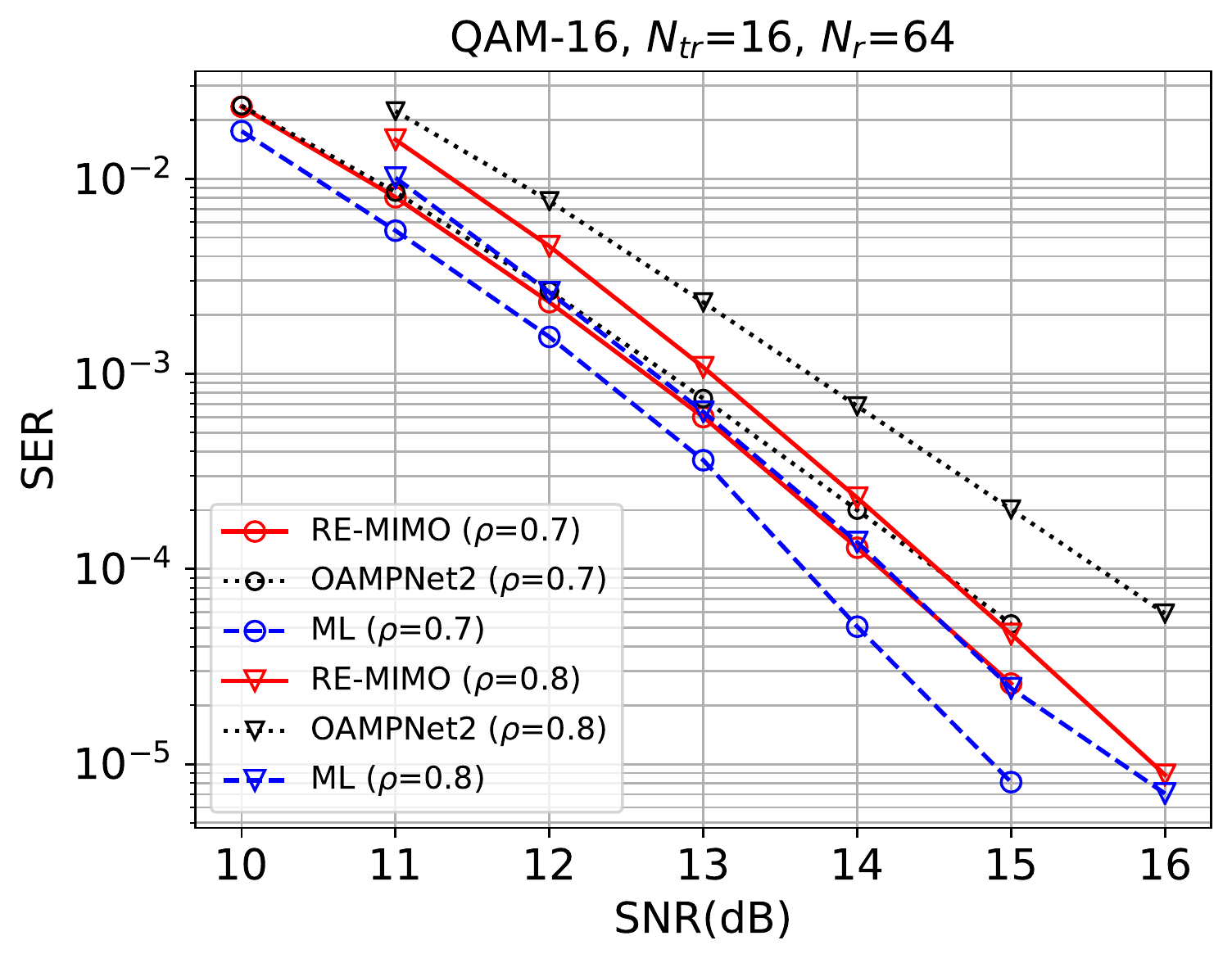}%
            \label{fig:corr_qam_64_NT_16}}
            \hfil
            \subfloat[Case II]{\includegraphics[width=3.2in]{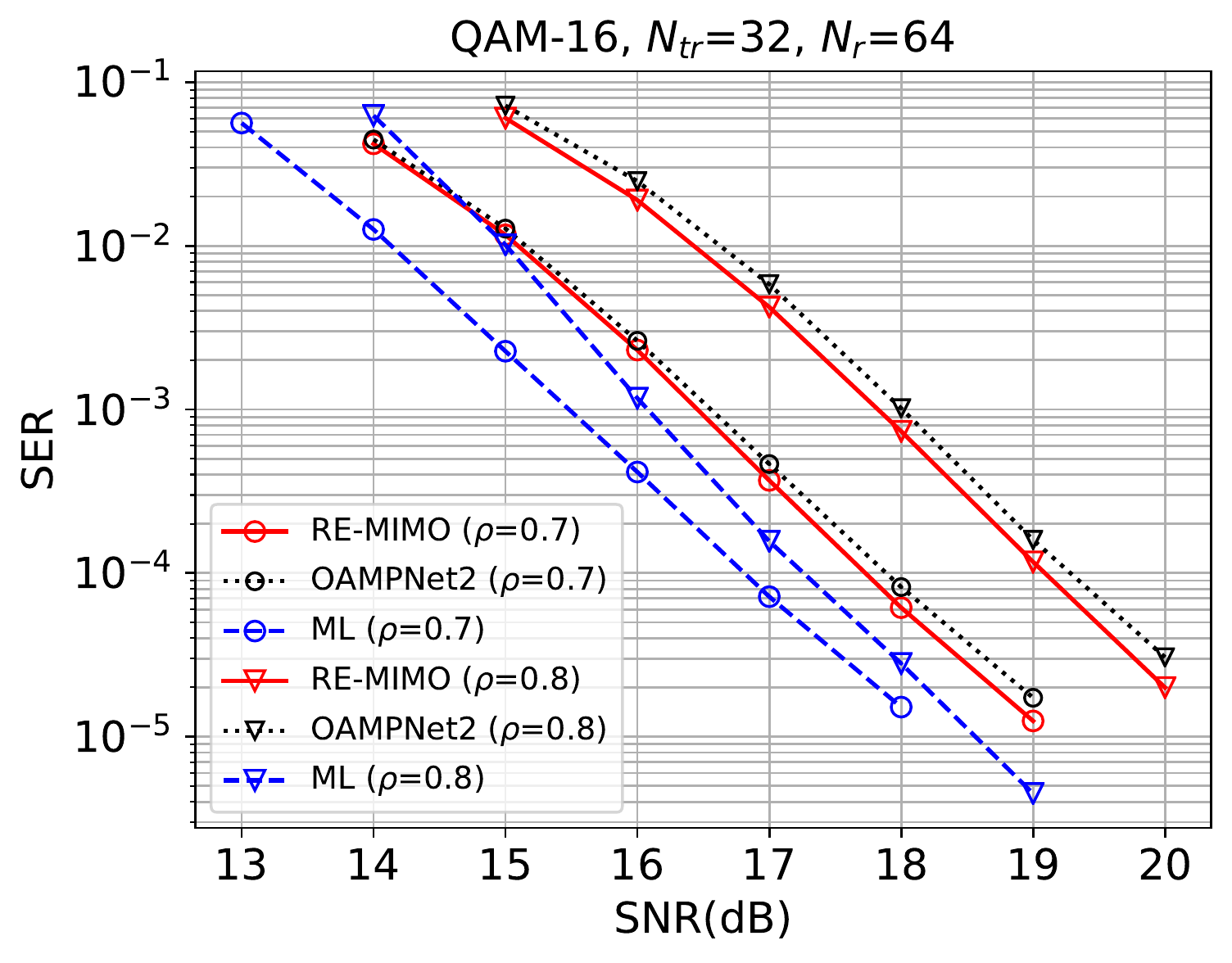}%
            \label{fig:corr_qam_64_NT_32}}
            \caption{SER vs. SNR for $16$, $32$ transmit antennas and $64$ receive antennas on partially correlated Rayleigh channels channels with QAM-16 modulation.}
            \label{fig:partial_corr_qam_16}
        \end{figure*}
        
        In practical scenarios, the correlation coefficient $(\rho)$ is unknown and can be assumed to lie in a range. Hence, we train the learning-based schemes on a range of $\rho$ values, i.e., $\rho \in [\rho_{min}, \rho_{max}]$. For correlated channels, we only consider OAMPNet-2 (baseline) and RE-MIMO from the previous experiments, as it is evident from sec.~\ref{sec:i.i.d} that all the other sophisticated detection schemes did not perform as well as RE-MIMO and OAMPNet-2. We train RE-MIMO with $16$ EP blocks and for each training iteration, we randomly sample a value of $\rho$ from a triangular probability distribution, $\rho \sim \mathcal{T}(a,b,c)$ with lower limit $a=\rho_{min}$, upper limit $b=\rho_{max}$, and the mode $c$ coinciding with the upper limit, i.e., $c=b$. We keep $\rho_{min}=0.55$, $\rho_{max}=0.75$ for correlated channels case and $\rho_{min}=0.65$, $\rho_{max}=0.85$ for partially correlated channels case.

        Fig.~\ref{fig:full_corr_qam_16} and Fig.~\ref{fig:partial_corr_qam_16} compare the SER performance for two system size ratios and two different values of $\rho$ for correlated and partially correlated channels respectively. Our single RE-MIMO outperforms the OAMPNet-2 which is separately trained for $N_{tr}=16$ and $N_{tr}=32$. Analysis of results for two different $\rho$ values suggests that increasing the $\rho$ value leads to an increase in the performance gap between RE-MIMO and the OAMPNet-2. This observation can be attributed to the fact that OAMPNet-2 makes a strong assumption on the channel matrix which leads to performance degradation in the case of realistic correlated channels. Unlike OAMPNet-2, RE-MIMO does not make any assumption on channel properties which makes it more robust and well suited for realistic channel realizations that deviate from i.i.d. Gaussian assumption.
        
    \subsection{Rayleigh MIMO channel with channel estimation errors}
        In this subsection, we evaluate the robustness of RE-MIMO to channel estimation errors. In most of the wireless communication systems, the receiver does not have access to perfect CSI. The receiver demodulates the received signal based on a noisy estimate of the true channel. To simulate the imperfect CSI scenario, we put an additional AWGN on the true channel matrix $\mathbf{H}$ before forwarding it to the receiver for demodulation. It can formally be stated as
        \begin{equation*}\label{eq:27} \tag{27}
            \mathbf{H}^{*} = \mathbf{H} + \mathbf{W}
        \end{equation*}
        where $\mathbf{H}^{*}$ represents the noisy estimate of the true channel matrix, $\mathbf{H}$ is the true channel matrix, and $\mathbf{W}$ is the noise representing the channel estimation error. Each element of $\mathbf{W}$ is sampled from a zero-mean i.i.d. Gaussian distribution, i.e. $w_{ij} \sim \mathcal{C} \mathcal{N}\left(0, \sigma^2_{w}\right)$, whose variance ($\sigma^2_{w}$) is related to the SNR as per the formula
        \begin{equation*}\label{eq:28} \tag{28}
            \text{SNR}=\frac{\mathbb{E}\left[||\mathbf{H}||_{\text{F}}^{2}\right]}{\mathbb{E}\left[||\mathbf{W}||_{\text{F}}^{2}\right]}
        \end{equation*}
        where $||\mathbf{H}||_{\text{F}}$ and $||\mathbf{W}||_{\text{F}}$ represents the Frobenius norm of the respective matrices.
        
        \begin{figure*}[!t]
            \centering
            \subfloat[Case I]{\includegraphics[width=3.2in]{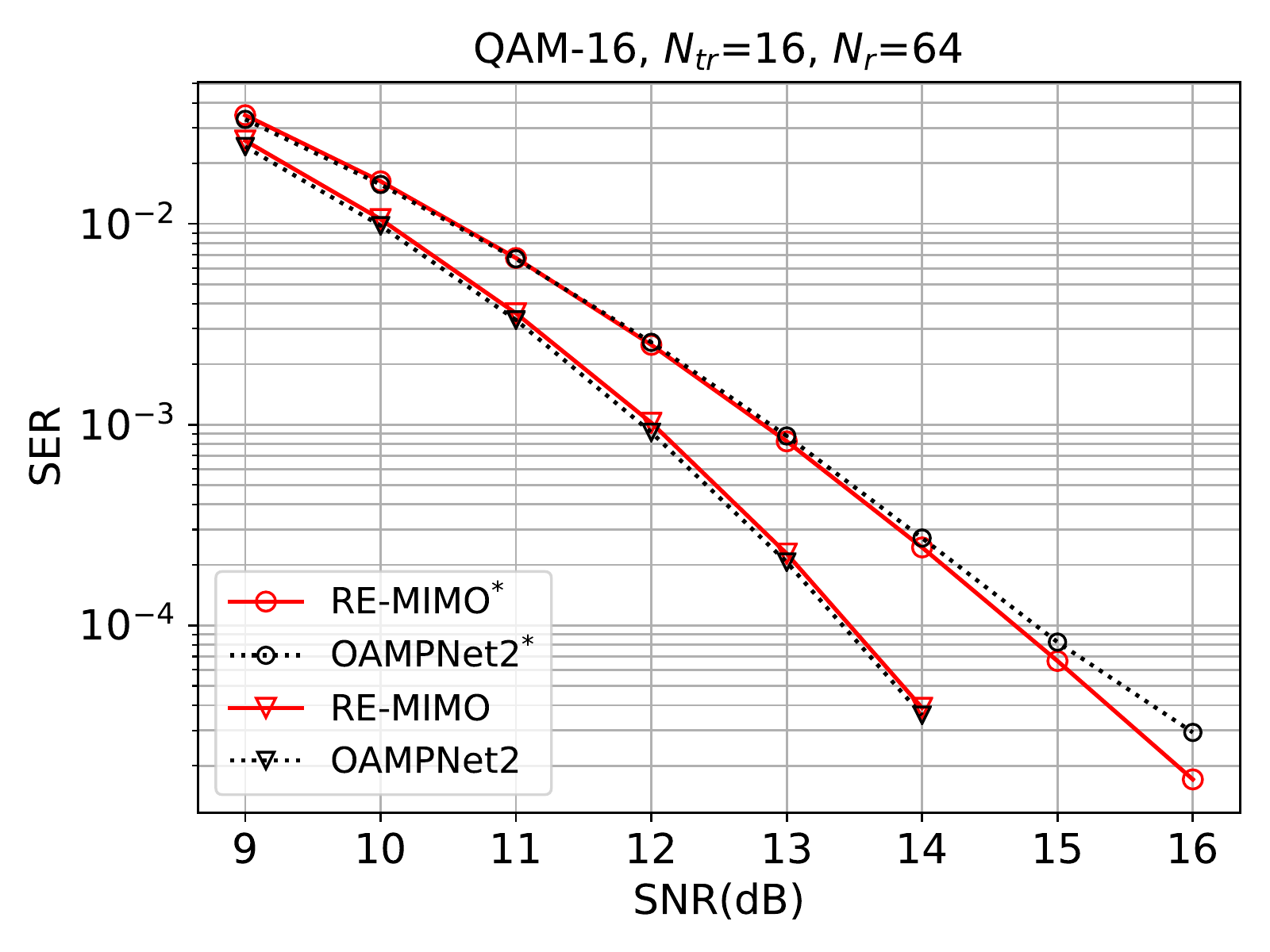}%
            \label{fig:chest_NT_16}}
            \hfil
            \subfloat[Case II]{\includegraphics[width=3.2in]{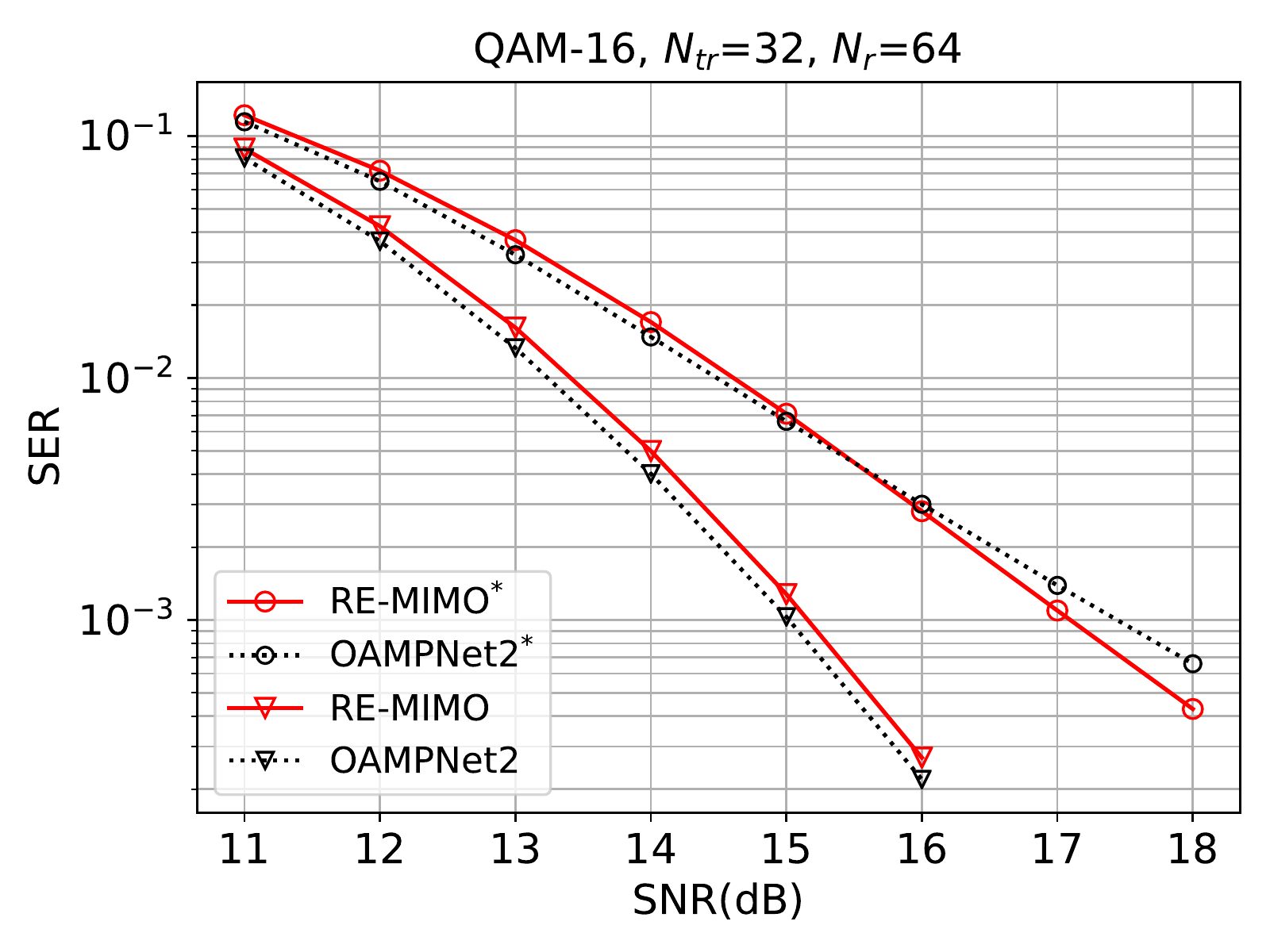}%
            \label{fig:chest_NT_32}}
            \caption{SER vs. SNR for $16$, $32$ transmit antennas, and $64$ receive antennas on i.i.d. Gaussian channels with QAM-16 modulation. RE-MIMO$^{*}$ and OAMPNet2$^{*}$ represent models trained on perfect CSI but tested on imperfect CSI while RE-MIMO and OAMPNet2 represent models trained and tested on perfect CSI.}
            \label{fig:chest_qam_16}
        \end{figure*}
        
        In our experiments, we set the SNR of channel estimation error noise at $20$ dB. To bring more perspective, we also include OAMPNet-2 in our experiments. We take the RE-MIMO and OAMPNet-2 models trained with perfect CSI on i.i.d. Gaussian channels from sec.~\ref{sec:i.i.d}, and test it on imperfect CSI, i.e., $\mathbf{H}^{*}$ is forwarded to these models instead of $\mathbf{H}$ for demodulation. We use an asterisk ($^*$) to differentiate between models with imperfect and perfect CSI, e.g., RE-MIMO$^*$ represents RE-MIMO trained on perfect CSI but tested on imperfect CSI while RE-MIMO represents RE-MIMO trained and tested on perfect CSI. Fig.~\ref{fig:chest_qam_16} depicts the performance of RE-MIMO and OAMPNet-2 in both the perfect and imperfect CSI scenarios for two system size ratios $0.25$ and $0.50$ on the QAM-16 modulation scheme. The performance of both the detectors degrades by almost the same amount when tested on imperfect CSI ($\mathbf{H}^{*}$) which suggests that RE-MIMO is as robust as other well established MIMO detectors.
        
    \subsection{Network dynamics}
        In this subsection, we investigate the limitations of RE-MIMO by training and testing it on different sets of user numbers. We also study the inner working dynamics of the attention mechanism in RE-MIMO.
        
        \subsubsection{Interpolation \& Extrapolation properties}
            To analyze the interpolation and extrapolation capabilities of RE-MIMO, we train it on every alternate value (even values) of $N_{tr}$ in the range $\left[16, 32\right]$, i.e., $N_{tr} \in \left\{n_{tr} \in \left[16, 32\right] : (\exists k \in \mathbb{Z})[n_{tr}=2k] \right\}$. We then test RE-MIMO on the values of $N_{tr}$ that lie in the aforementioned range but were left out during training (odd values), to gauge the interpolation capability of RE-MIMO. For analyzing the extrapolation capability, we test RE-MIMO on values of $N_{tr}$ beyond the training range, i.e., $N_{tr} \in \{36,40\}$.
            
\begin{figure*}[!t]
            \centering
            \subfloat[Case I]{\includegraphics[width=2.3in]{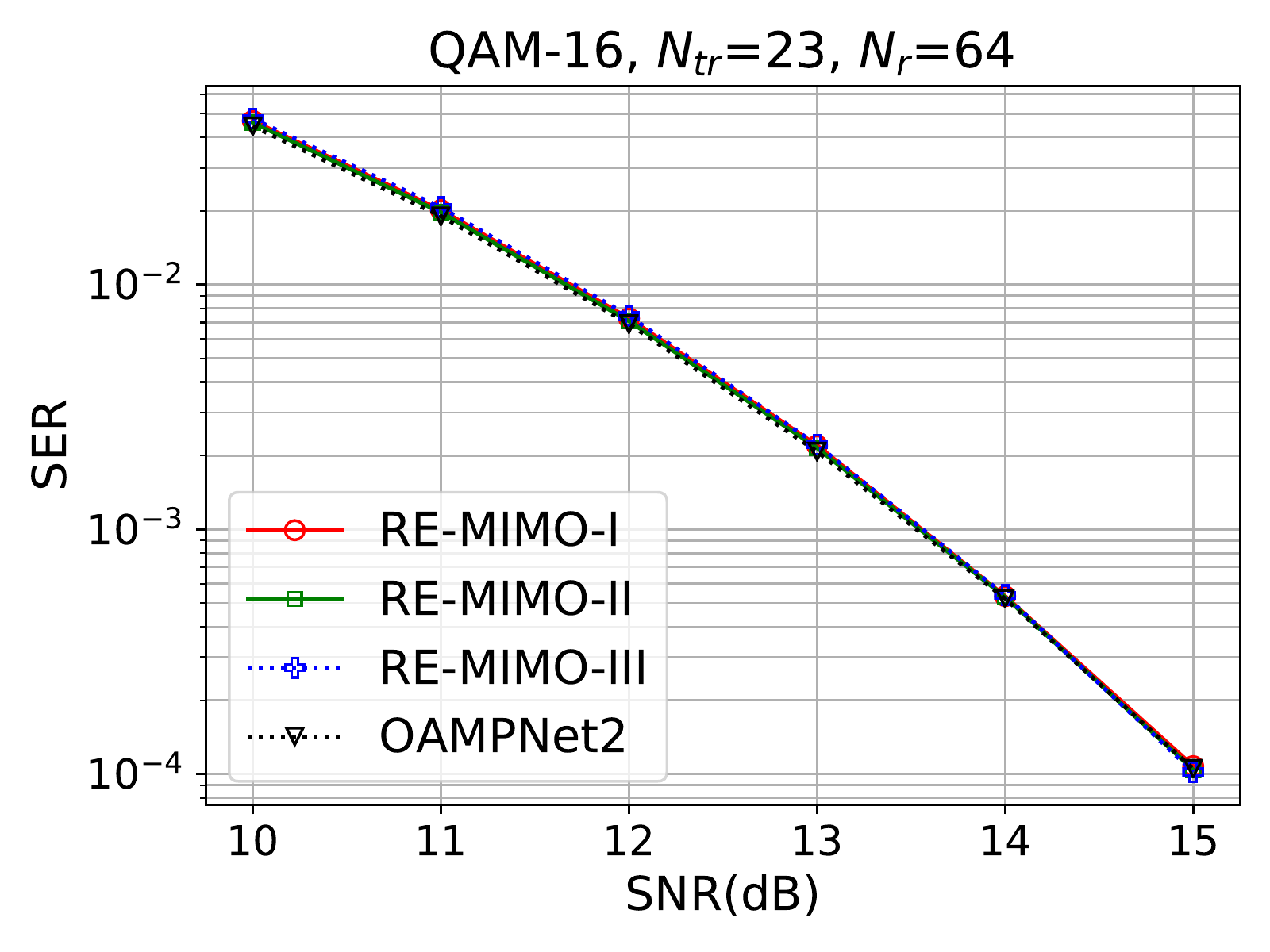}%
            }
            \hfil
            \subfloat[Case II]{\includegraphics[width=2.3in]{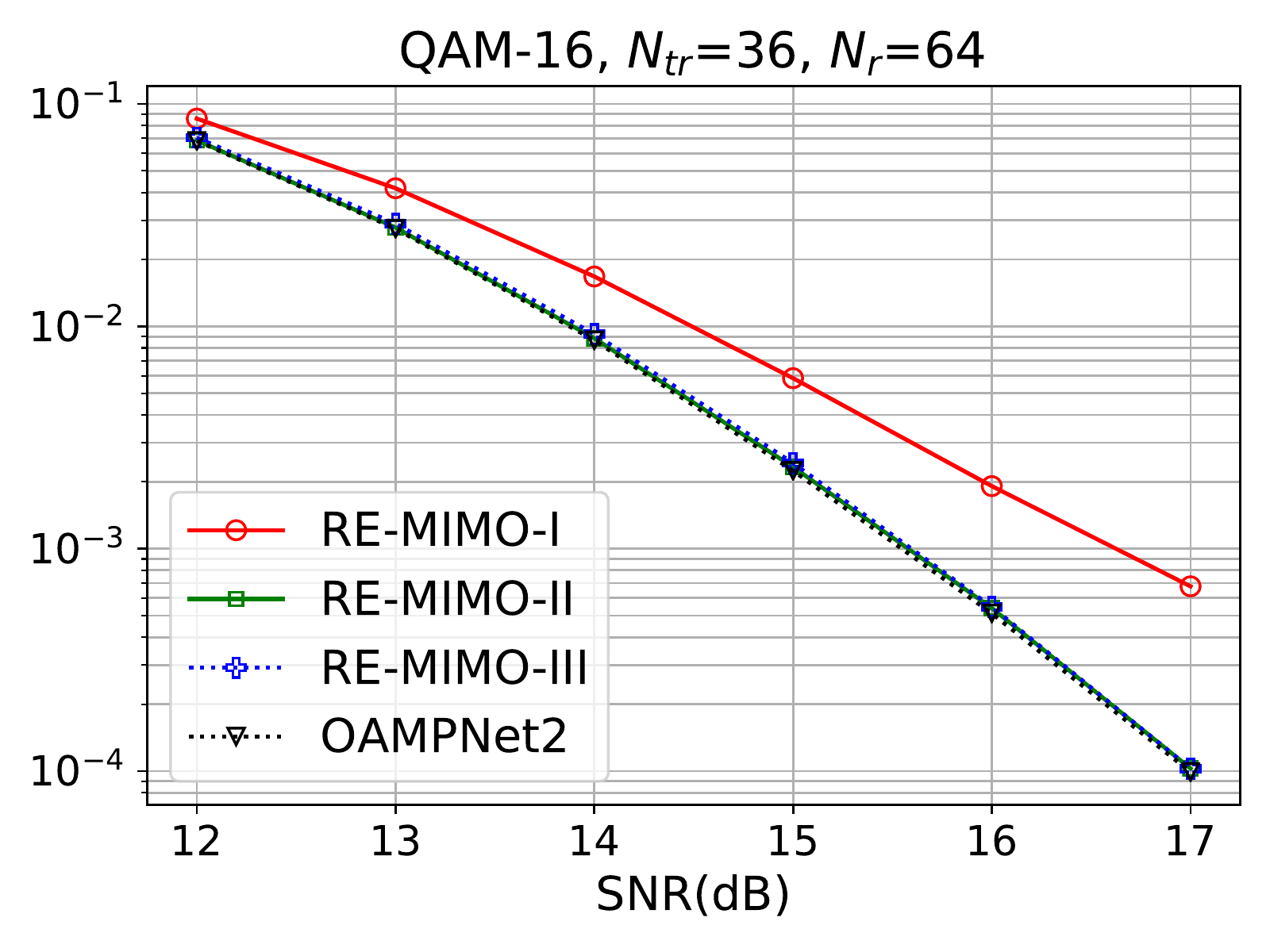}%
            }
            \hfil
            \subfloat[Case III]{\includegraphics[width=2.3in]{images/interpol_NT_36_NR64.pdf}%
            }
            \caption{SER vs. SNR plots illustrating the interpolation and extrapolation results of RE-MIMO on i.i.d. Gaussian channels with QAM-64 modulation.}
            \label{fig:interpolation}
        \end{figure*}

            Fig.~\ref{fig:interpolation} depicts the SER performance of RE-MIMO for MIMO systems with $N_{tr} \in \{23,36,40\}$ on QAM-16 modulation and i.i.d. Gaussian channel matrix. We compare the performance of the above explained version of our detector (RE-MIMO-I) with OAMPNet-2 (trained separately for each $N_{tr}$ value), a full version of our detector (RE-MIMO-II) which is trained on all of the values of $N_{tr}$ in the range $\left[16, 44\right]$ without skipping, and an exclusive version of our detector (RE-MIMO-III) where we train a separate detector for each value of $N_{tr} \in \{23,36,40\}$.
            
            It is evident from Fig.~\ref{fig:interpolation} that RE-MIMO-I elegantly interpolates to intermediate (odd) values of $N_{tr}$ without any performance degradation. This suggests that the proposed RE-MIMO is scalable to MIMO systems with higher configurations in the sense that in case of higher values of $N_{r}$, training the RE-MIMO on every alternate (or possibly more sparse) values of $N_{tr}$ in the required range suffices for the entire range of users. Another very important and interesting observation we make is that training RE-MIMO to attend to a variable number of users does not lead to any performance degradation as compared to exclusively training of RE-MIMO for a fixed number of users. On the other hand, the performance of RE-MIMO degrades significantly when extrapolated to $N_{tr}$ values far beyond the training range, which we expect to be a rare occurrence for a properly provisioned system.
        
        \subsubsection{Attention Interpretation}
            \begin{figure*}[!t]
                \centering
                \subfloat[Case I]{\includegraphics[width=2.7in]{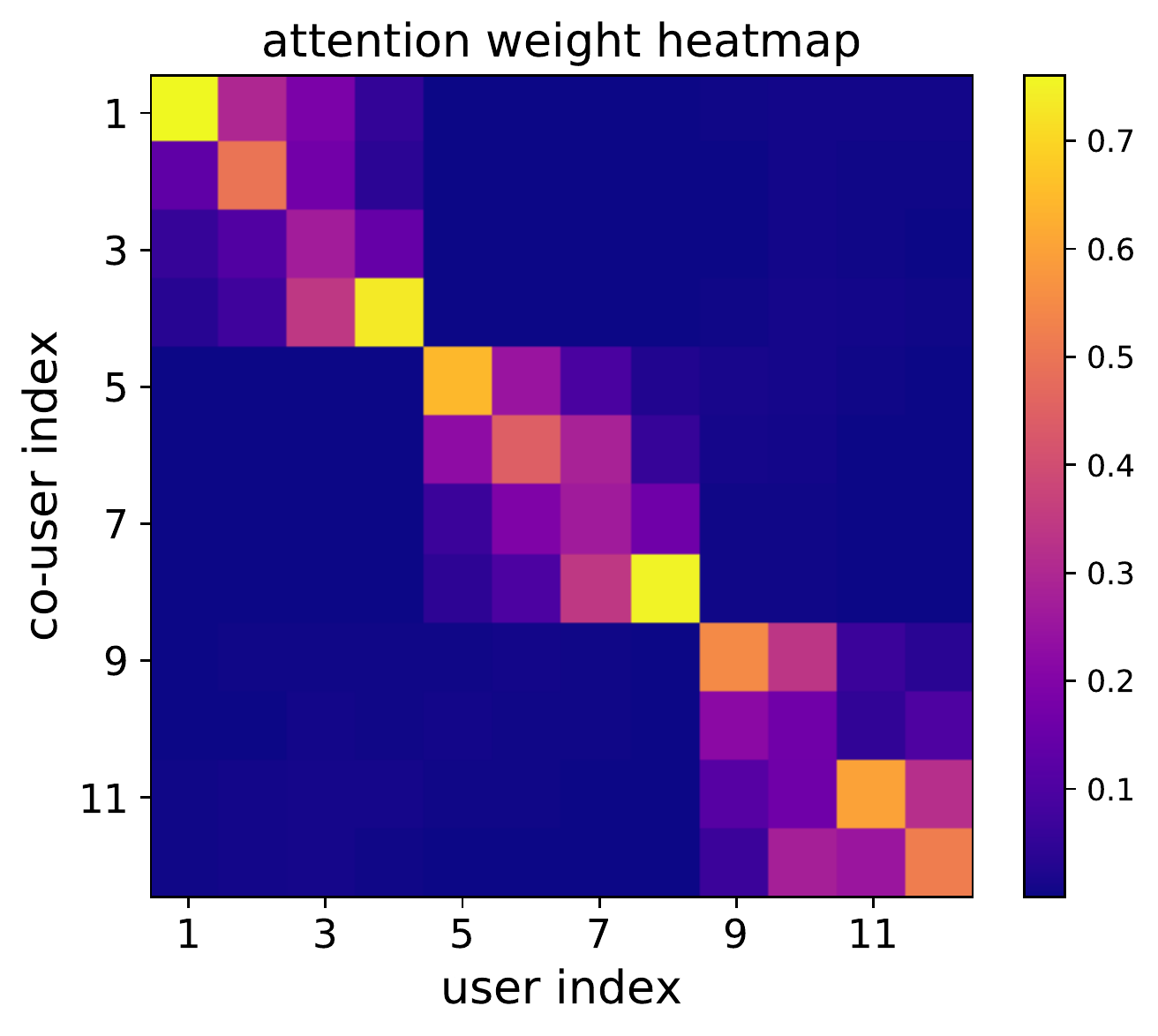}%
                }
                \hfil
                \subfloat[Case II]{\includegraphics[width=2.7in]{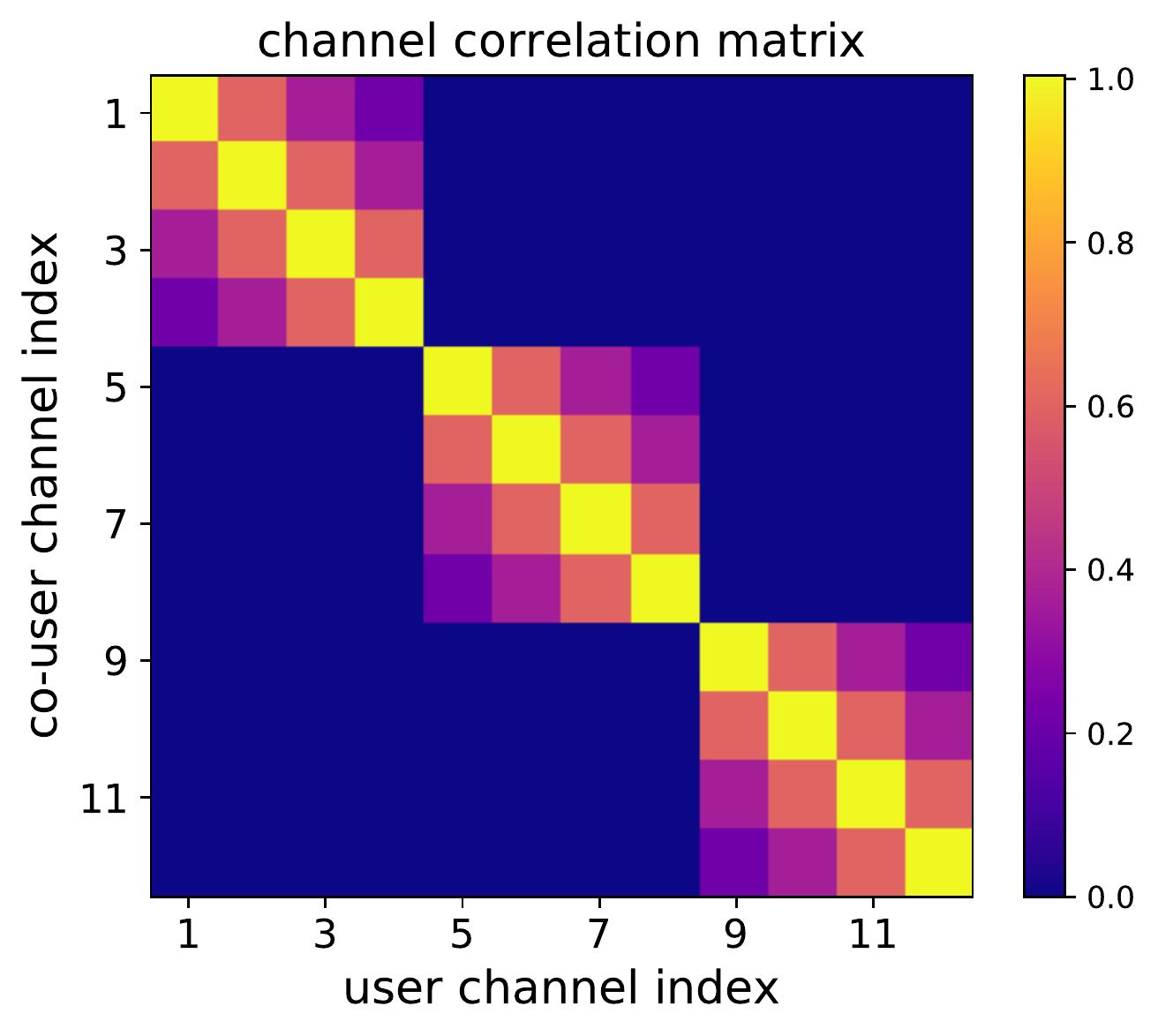}%
                }
                \caption{The RE-MIMO evidently learns to figure out to pay attention only to the users belonging to the non-orthogonal (parent) set of co-transmitters and ignores the remaining co-transmitters with orthogonal channels. The resemblance between the attention heatmap and channel alignment suggests a possible interpretation of the RE-MIMO attention mechanism as a neural interference cancellation mechanism for MIMO detection.}
                \label{fig:attention_heatmap}
            \end{figure*}
            
            It is a known fact that mutual interference between co-transmitters depends on their channel alignment. This suggests that the attention $(\alpha_{ij})$ RE-MIMO pays to user $j$ while decoding the symbol for user $i$, should be related to the alignment between $\mathbf{h}_{i}$ and $\mathbf{h}_{j}$. We conduct a controlled experiment to verify that our notion of attention and interference align with each other. We train a simplified version of RE-MIMO with eight EP blocks, where each EP block has only one attention head, for a MIMO system with $N_{r}=32$ and a fixed value of $N_{tr}=12$. For each training sample, we begin with $12$ orthogonal channel vectors. Then we proceed to form three sets of channels (users) containing $4$ channels each, i.e., $\mathcal{A}=\{\mathbf{h}_{i}\}_{i=1}^{4}, \mathcal{B}=\{\mathbf{h}_{i}\}_{i=5}^{8}, \mathcal{C}=\{\mathbf{h}_{i}\}_{i=9}^{12}$. Subsequently, we introduce correlation among each channel set separately by post-multiplying it with a mixing matrix such that the resulting channels belonging to one set are a linear combination of the original orthogonal channels of that set. Henceforth, channels belonging to the same set no longer remain orthogonal to each other but will remain orthogonal with every other channel outside its parent set.
            
            This is a controlled experiment in the sense that we control which user will interfere with whom. We train RE-MIMO on such a controlled dataset for QAM-16 modulation. Fig.~\ref{fig:attention_heatmap} illustrates the attention heat map obtained from the self-attention layer of the third EP block. Our RE-MIMO successfully figures out the set of users it should pay attention to while decoding a particular user symbol.
        
    \subsection{Training}
        In this subsection, we discuss the methodology used for training the learning-based algorithms for our experiments.
        
        The learning-based schemes used in our experiments, i.e., RE-MIMO, DetNet, OAMPNet, and OAMPNet-2, were implemented in PyTorch~\cite{pytorch}. We frame the MIMO problem to its equivalent real-valued representation for PyTorch implementation. We keep the training batch size for OAMPNet/OAMPNet-2 and DetNet at $5$K and $1$K samples respectively. OAMPNet, OAMPNet-2, and DetNet are trained with specifications as mentioned in their respective papers and we make sure to train the models until the SER performance stops improving. For each combination of $\{N_{r}, N_{tr}, \text{QAM-order}\}$, we initialize and train a separate network for OAMPNet, OAMPNet-2, and DetNet, each of which is trained over a range of SNR values. During training, we target the SNR range to achieve an SER of $10^{-3}-10^{-2}$, as most of the error correcting schemes operate in this range.
        
        \begin{figure*}[!t]
            \centering
            \subfloat[Case I]{\includegraphics[width=2.7in]{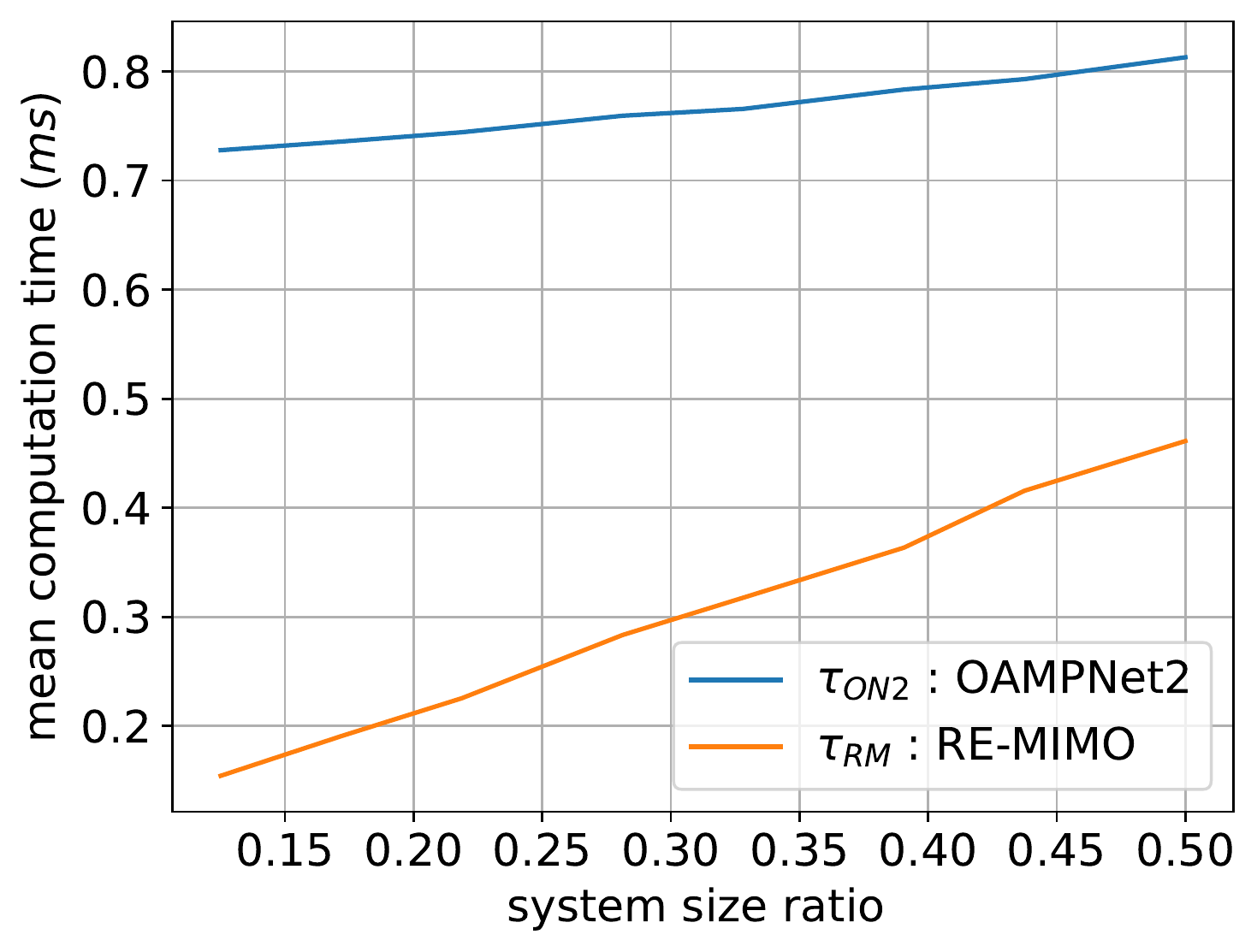}%
            }
            \hfil
            \subfloat[Case II]{\includegraphics[width=2.7in]{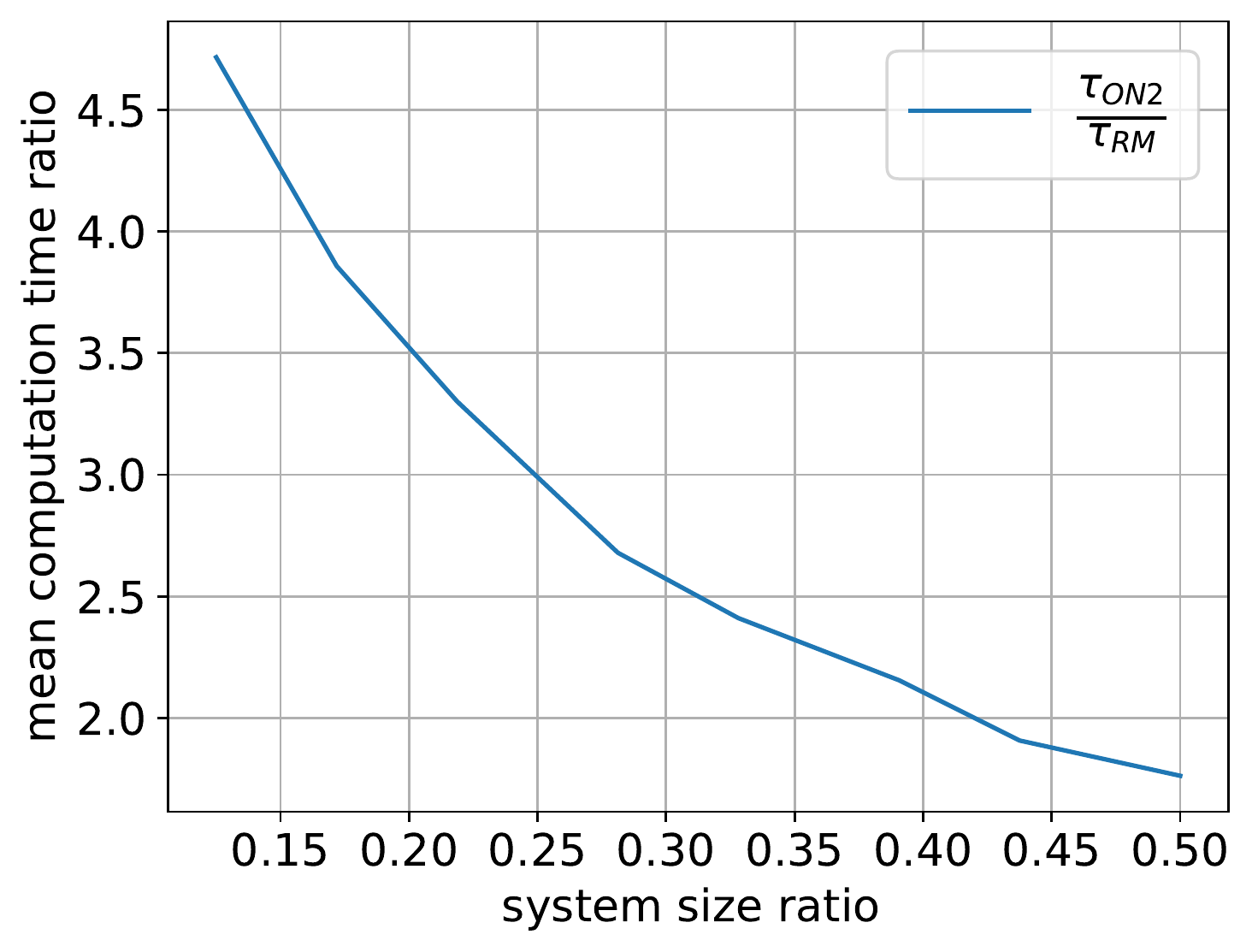}%
            }
            \caption{Plots depicting the mean computation time and its ratio for OAMPNet2 and RE-MIMO detectors.}
            \label{fig:mct}
        \end{figure*}
        
        For a given combination of $\{N_{r}, \text{QAM-order}\}$, we train a single RE-MIMO for all the number of transmitters in the range $N_{tr} \in \left[{N_r}/{4}, {N_r}/{2}\right]$. We use Adam optimizer~\cite{adam} with an initial learning rate of $10^{-4}$ for training RE-MIMO. We deploy a learning rate scheduler (PyTorch \textit{ReduceLROnPlateau}) to dynamically adjust the learning rate based on validation set measurements. The loss function in (\ref{eq:11}) is used to train the network with an equal weighing of loss at each iteration step, i.e., $w_t=1/T$. We train RE-MIMO for $120$ epochs with a mini-batch size of $256$ samples, and after each epoch ($5$K iterations), we compute the loss on the validation set and if the loss hits a plateau, i.e., the performance metric stopped improving, the learning rate is reduced by a factor $k=0.91$. Each training iteration of RE-MIMO involves sampling a value of $N_{tr}$ from a triangular probability distribution, $N_{tr} \sim \mathcal{T}(a,b,c)$, with lower limit $a=N_r/4$, upper limit $b=N_r/2$, and the mode $c$ coinciding with the upper limit, i.e., $c=b$. Interference is more when $N_{tr}$ is high and hence, the detector needs to be trained more often on higher values of $N_{tr}$. A target SNR range is determined for the sampled value of $N_{tr}$ and subsequently, all the samples in the training mini-batch have the same $N_{tr}$ with their SNR values uniformly distributed on $\mathcal{U} \left(\text{SNR}_{min}, \text{SNR}_{max}\right)$. The validation set contains samples pertaining to two different system sizes, $\{N_{tr}=16, N_{r}=64\}$ and $\{N_{tr}=32, N_{r}=64\}$. Each system size has a different target SNR range. The validation set contains $5$k samples for each SNR value in the target SNR range of the respective system size. The final validation loss is the mean of losses over the entire validation set samples of both the system sizes. After training, the final evaluation of RE-MIMO is done on the test set consisting of $10^7$ samples for each SNR value.
        
    \subsection{Computational complexity}
        Symbol detection takes $\mathcal{O}(N_{r}^2 N_{tr}+ N_rN_{tr}^{2})$ per EP block in RE-MIMO. To bring more viewpoints, AMP has a complexity of $\mathcal{O}(N_{tr}N_{r})$ per iteration step, which is ascribed to the multiplication of the channel matrix with the residual vector. The MMSE detector computes a pseudo inverse for inverting the signal, owing it a complexity of $\mathcal{O}(N_{tr}^{3}+N_{r}N_{tr}^{2})$. Similarly, OAMPNet-2 also requires matrix pseudo inversion at each iteration step, causing a per layer complexity of $\mathcal{O}(N_{r}^3+ N_{tr}^{3}+ N_{r}N_{tr}^{2}+N_{tr}N_r^{2})$. For RE-MIMO, $N_{r}^2 N_{tr}$ dominates the complexity metric, that grows linearly with $N_{tr}$. Hence, the number of computations required is less when a smaller number of users are connected to the BS. On the other hand, the dominating term for the OAMPNet-2 detector is $N_r^3$, which does not depend on $N_{tr}$, and hence, the detection complexity remains high even in the case with fewer operational users. Moreover, RE-MIMO can leverage the benefits of the extensive parallelization property of the Transformer network.
        
        Moving beyond $\mathcal{O(.)}$ analysis, we investigate the mean computation time (mct) taken by RE-MIMO and OAMPNet-2 to decode $\hat{\mathbf{x}}$ given $\{\mathbf{y, H, \sigma}\}$, when simulated on the similar hardware platform (NVIDIA TITAN V GPU). For this purpose, we take the already trained models of RE-MIMO and OAMPNet-2 for i.i.d. Gaussian channels with QAM-16 modulation from~\ref{sec:i.i.d} and analyze the computational time taken by both of these models during the inference phase. The findings in Fig.~\ref{fig:mct} for $N_r = 64$ are in complete agreement with our earlier postulate. We observe that the mct for RE-MIMO ($\tau_{\text{RM}}$) is lower in the case of lower system size ratios and increases linearly with $N_{tr}$. While for the OAMPNet-2, the mct ($\tau_{ON}$) remains high even for the lower system size ratios and does not vary much with $N_{tr}$.
        
\section{Conclusion}
\label{sec:conclusion}
    In this work, we incorporate several properties that make neural MIMO receivers more practical. We present a novel neural MIMO receiver that detects symbols transmitted from a varying number of users. Possible future directions of research include the development of RE-MIMO extensions capable of handling multiple modulation schemes and incorporating imperfect CSI, more realistic channels and noise models into RE-MIMO design.
    
    RE-MIMO offers a receiver design solution in the form of a single permutation equivariant neural detector for all the possible number of transmitters. RE-MIMO achieves this by sharing the same set of parameters with all the users. The benefits of learning a single set of parameters are threefold; 1) fewer parameters to train, 2) the network can be extended to possibly any number of users, and 3) protection against overfitting by avoiding learning spurious patterns based on an artificial ordering of the inputs. RE-MIMO incorporates generative modeling wrapped around by a discriminative network which helps the network achieve state-of-the-art SER performance on correlated channels.
        

%





\ifCLASSOPTIONcaptionsoff
  \newpage
\fi



\bibliographystyle{IEEEtran}
\bibliography{IEEEabrv, references.bib}
%



%

\begin{IEEEbiographynophoto}{Kumar Pratik}
received the B.Tech. degree with majors in Engineering Physics and a minor in Electrical Engineering from the Indian Institute of Technology Bombay, India, in 2018. He graduated with M.S. degree in Artificial Intelligence from the University of Amsterdam, The Netherlands, in 2020. His interests are in the areas of machine learning, computer vision, and statistics. He is currently a Deep Learning Research Engineer at Qualcomm Research AI, Amsterdam.
\end{IEEEbiographynophoto}

\begin{IEEEbiographynophoto}{Bhaskar D. Rao}
(S'80–M'83–SM'91–F'00) is currently
a Distinguished Professor in the Department of
Electrical and Computer Engineering and the holder
of the Ericsson Endowed Chair in wireless access
networks at the University of California, San Diego, CA, USA. He received the 2016 IEEE Signal Processing Society Technical Achievement Award. His research interests include digital signal processing,
estimation theory, and optimization theory, with applications
to digital communications, speech signal
processing, and biomedical signal processing.
\end{IEEEbiographynophoto}

\begin{IEEEbiographynophoto}{Max Welling} is a research chair in Machine Learning at the University of Amsterdam and a VP Technologies at Qualcomm. He has a secondary appointment as a senior fellow at the Canadian Institute for Advanced Research (CIFAR). He has served as associate editor in chief of IEEE TPAMI from 2011-2015. He serves on the board of the NIPS foundation since 2015 and has been program chair and general chair of NeurIPS in 2013 and 2014 respectively. He was also program chair of AISTATS in 2009 and ECCV in 2016 and general chair of MIDL 2018. He is recipient of the ECCV Koenderink Prize in 2010. Welling is co-founder and board member of the Innovation Center for AI (ICAI) and the European Lab for Learning and Intelligent Systems (ELLIS). He directs the Amsterdam Machine Learning Lab (AMLAB), and co-directs the Qualcomm-UvA deep learning lab (QUVA) and the Bosch-UvA Deep Learning lab (DELTA). 
\end{IEEEbiographynophoto}







\end{document}